\documentclass[aps,prx,preprint,onecolumn,citeautoscript,footinbib,eqsecnum,superscriptaddress]{revtex4-2}
\usepackage{feynmp}
\usepackage[section]{placeins}
\usepackage{subcaption}
\usepackage{amsmath}
\usepackage{amssymb}
\usepackage{bbm}
\usepackage{braket}
\usepackage{xcolor}
\usepackage{pifont}
\usepackage{slashed}
\usepackage[mathscr]{euscript}
\usepackage[shortlabels]{enumitem}
\usepackage{tikz-feynman}
\usepackage[papersize={8.5in,11in}]{geometry}

\allowdisplaybreaks

\usepackage{graphicx}
\usepackage{subcaption}
\usepackage[colorlinks=true]{hyperref}
\usepackage{comment}

\newcommand{\joerg}[1]{{\color{black}{ #1}}}

\renewcommand{\approx}{\simeq}
\renewcommand{\Re}{\text{Re}}
\renewcommand{\Im}{\text{Im}}

\definecolor{wrongultramarine}{rgb}{1,0.5,0}

\newcommand{\rd}{{\rm d}}
\newcommand{\sgn}{{\rm sgn\,}}

\newcommand{\calN}{{\mathcal N}}

\newcommand{\calF}{{\mathcal F}}
\newcommand{\calD}{{\mathcal D}}
\hypersetup{
    bookmarks=true,         
    unicode=false,          
    pdftoolbar=true,        
    pdfmenubar=true,        
    pdffitwindow=false,     
    pdfstartview={FitH},    
    pdfsubject={},   
    pdfcreator={},   
    pdfproducer={}, 
    pdfkeywords={} {} {}, 
    pdfnewwindow=true,      
    colorlinks=true,       
    linkcolor=blue, 
    citecolor=blue,        
    filecolor=magenta,      
    urlcolor=blue           
}

\tikzset{
  mid arrow/.style={postaction={decorate,decoration={
        markings,
        mark=at position .575 with {\arrow[#1]{stealth}}
      }}},
  near arrow/.style={postaction={decorate,decoration={
        markings,
        mark=at position .275 with {\arrow[#1]{stealth}}
      }}},
   far arrow/.style={postaction={decorate,decoration={
        markings,
        mark=at position .800 with {\arrow[#1]{stealth}}
      }}},
}

\begin{document}
\preprint{\href{https://arxiv.org/abs/2308.01956}{arXiv:2308.01956}}

\title{Cyclotron resonance and quantum oscillations\\ of critical Fermi surfaces}
\begin{abstract}
Kohn's theorem places strong constraints on the cyclotron response of Fermi liquids. Recent observations of a doping dependence in the cyclotron mass of La$_{2-x}$Sr$_x$CuO$_4$ (Legros {\it et al.}, Phys. Rev. B {\bf 106}, 195110 (2022))
are therefore surprising because the cyclotron mass can only be renormalized by large momentum umklapp interactions which are not expected to vary significantly with doping. We show that a version of Kohn's theorem continues to apply to disorder-free non-Fermi liquids with a critical boson near zero momentum. However, marginal Fermi liquids arising from a spatially random Yukawa coupling between the electrons and bosons do give rise to significant corrections to the cyclotron mass which we compute. This is the same theory which yields linear-in-temperature resistivity and other properties of strange metals at zero fields (Patel {\it et al.}, Science {\bf 381}, 790 (2023)).
\end{abstract}

\author{Haoyu Guo}
\affiliation{Laboratory of Atomic and Solid State Physics, Cornell University,
142 Sciences Drive, Ithaca NY 14853-2501, USA}
\affiliation{Department of Physics, Harvard University, Cambridge MA 02138, USA}

\author{Davide Valentinis}
\affiliation{Institut f\"ur Quantenmaterialien und Technologien, Karlsruher Institut
f\"ur Technologie, 76131 Karlsruhe, Germany}
\affiliation{Institut f\"ur Theorie der Kondensierten Materie, Karlsruher Institut
f\"ur Technologie, 76131 Karlsruhe, Germany}

\author{J\"org Schmalian}
\affiliation{Institut f\"ur Theorie der Kondensierten Materie, Karlsruher Institut
f\"ur Technologie, 76131 Karlsruhe, Germany}
\affiliation{Institut f\"ur Quantenmaterialien und Technologien, Karlsruher Institut
f\"ur Technologie, 76131 Karlsruhe, Germany}

\author{Subir Sachdev}
\affiliation{Department of Physics, Harvard University, Cambridge MA 02138, USA}

\author{Aavishkar A. Patel}
\affiliation{Center for Computational Quantum Physics, Flatiron Institute, New York,
New York, 10010, USA}

\date{\today}

\maketitle
\newpage

\tableofcontents

\section{Introduction}

In modern condensed matter physics, magnetic field responses have become the go-to tool for probing Fermi surfaces (FS) of conventional Fermi liquids (FL) \cite{Ashcroft1976}. In this work, we explore the possibility of utilizing such responses to uncover novel physics beyond the Landau Fermi liquid theory, namely non-Fermi liquids (NFL) and marginal Fermi liquids (MFL).

The first probe we discuss is the cyclotron resonance, which refers to the collective motion of the electrons under an external magnetic field. The collective motion can be measured through optical conductivity which reflects as a resonance at the cyclotron frequency $\omega_c^*={eB}/{m_c}$, where $m_c$ is the cyclotron mass. Legros {\it et.al.} \cite{Legros2022}
have recently studied the cyclotron resonance of the cuprate compound La$_{2-x}$Sr$_x$CuO$_4$ using time-domain Thz spectroscopy. They observed that the cyclotron mass in cuprates is renormalized away from the band mass as a function of doping. This observation is in contrast with  Kohn's theorem \cite{Kohn1961,Kanki1997} stating that the cyclotron frequency of a translational and Galilean invariant electron gas is not altered by electron-electron interaction. Therefore, a candidate theory to explain the experiment should break at least one of these two symmetries. In Ref.~\cite{Kanki1997}, the effect of umklapp scattering was shown to be nonzero but relatively small, so we expect the renormalization of the cyclotron mass to be due to disorder effects. In this manuscript, we examine the effect of the usual potential disorder and the more novel interaction disorder \cite{Patel2023} which gave rise to a universal mechanism for strange metallic behaviors. We find that the interaction disorder can more effectively renormalize the cyclotron mass and produce a sharper resonance peak.

The second probe we discussed in this paper is quantum oscillation, referring to the phenomenon that various physical quantities oscillate as a function of $1/B$ with a period set by the Fermi surface cross section. 
We propose that certain non-Fermi liquid or marginal-Fermi liquid aspects can be extracted from the local density states $\rho_\text{loc}(\epsilon,B)$, where $\epsilon$ is the energy. First, while the $1/B$ oscillation in $\rho_\text{loc}(\epsilon,B)$ still has a period set by the Fermi surface cross section, the Dingle factor as a function of $\epsilon$ contains information about the single-particle scattering rate, and so directly yields information about NFL/MFL aspects. Second, by measuring $\rho_\text{loc}(\epsilon,B)$ at fixed $B$ and scanning $\epsilon$, one can observe an oscillatory response in $\epsilon$. In a Fermi liquid, the period of such a oscillation is given by the bare cyclotron frequency $\omega_c$, but in a NFL or MFL, the oscillation becomes aperiodic in the sense that the period grows at higher $\epsilon$.

To illustrate our proposal, we will consider two types of theoretical models based on the previously developed large $N$ formalism \cite{Esterlis2021,Guo2022,Patel2023} of strongly interacting Fermi surface: the $g$-model and the $g'$-model. The $g$-model \cite{Esterlis2021}, inspired by Sachdev-Ye-Kitaev models \cite{Sachdev1993,kitaev2015talk,Sachdev2015,Maldacena2016e},  consists of $N\to\infty$ flavors of complex fermions $\psi_i$ with identical Fermi surfaces coupled to $N$ flavors of gapless real boson $\phi_l$ through random Yukawa couplings in flavors. Despite the randomness in flavor indices, the model is translational invariant and its large-$N$ saddle point describes a non-Fermi liquid similar to those obtained from Eliashberg equations \cite{PALee89,Polchinski:1993ii}. The $g'$-model \cite{Aldape2022,Esterlis2021,Patel2023} can be obtained from the $g$-model by making the Yukawa couplings random in space as well. The large $N$ saddle point of the $g'$ model describes a marginal Fermi liquid with strange metal behavior, including the linear-in-temperature resistivity.

The rest of the paper is organized as follows:

In Sec.~\ref{sec:gpmodel}, we investigate the $g'$-model which contains interaction disorder and is technically simpler. We review its formulation in the large $N$ limit and analyze the saddle point equations in a uniform magnetic field both analytically and numerically. Next we discuss its quantum oscillation responses in the local density of states and the de Haas-van Alphen effect. To conclude the section we compute the optical conductivity of the model and discuss the cyclotron resonance.

In Sec.~\ref{sec:gmodel}, we performed a parallel analysis of the translational invariant $g$-model. In particular, we show that optical conductivity of the model is exactly of the Drude form, re-establishing Kohn's theorem in the context of NFLs.

\section{Disordered $g'$-Model}\label{sec:gpmodel}
\FloatBarrier
\subsection{Action}

We review the $g'$-model in 2+1D spacetime which has appeared in previous works \cite{Esterlis2021,Patel2023}. The model consists of $N$ flavors of complex fermions $\psi_i$ and real bosons $\phi_l$, coupled through a Yukawa coupling random in both flavor and space. The action of the model is
\begin{equation}\label{eq:Sgprime}
\begin{split}
   S =& \int \rd \tau \sum_{\vec{k}} \sum_{i=1}^{N}\psi_{i\vec{k}}^\dagger(\tau)\left[\partial_\tau+\varepsilon_{\vec{k}+\vec{A}}-\mu\right]\psi_{i\vec{k}}(\tau) \\
      +\frac{1}{2}&\int \rd \tau \sum_{i=1}^{N}\sum_{\vec{q}} \phi_{i,\vec{q}} \left[-\partial_\tau^2+\omega_{\vec{q}}^2+m_b^2\right]\phi_{i,-\vec{q}}(\tau)\\
      +\frac{1}{N}&\int\rd^2 \vec{r}\rd \tau g'_{ijl}(\vec{r})\psi_i^\dagger(\vec{r},\tau)\psi_{j}(\vec{r},\tau)\phi_l(\vec{r},\tau)\,.
\end{split}
\end{equation} In Eq.\eqref{eq:Sgprime}, $\varepsilon_{\vec{k}+\vec{A}}$ is the fermion dispersion which we take to be $\varepsilon_{\vec{k}}=\vec{k}^2/(2m)$. $\vec{A}(x)=(-eBx_2,0,0)$ is the vector potential where $B$ is the magnetic field.   The boson dispersion is $\omega_{\vec{q}}^2=\vec{q}^2$ where we have set the `velocity of light' to unity. The bosons also posses a bare mass $m_b^2$ which will be tuned to criticality (There is some subtlety in the approach to criticality due to irrelevant operators \cite{Esterlis2021}, but we focus only on the critical point near zero temperature and ignore these issues).
The Yukawa couplings $g'$ are Gaussian spatially random variables satisfying
\begin{equation}\label{}
  \overline{g'_{ijl}(\vec{r})}=0\,,\qquad \overline{g'^*_{ijl}(\vec{r})g'_{abc}(\vec{r}')}=g'^2\delta^2(\vec{r}-\vec{r}')\delta_{ia}\delta_{jb}\delta_{lc}\,.
\end{equation}
As we are interested in the properties of the normal phase, we have allowed the couplings to be complex to suppress superconductivity. For theoretical simplicity we don't include any potential disorder, but we will comment on its effect later.

We take the $N\to\infty$ limit, and average over disorder configurations (at leading large $N$ order there is no difference between anneal or quench average \cite{Chowdhury2018}) to obtain the $G$-$\Sigma$ action:
\begin{equation}\label{eq:GSprime}
  \begin{split}
     \frac{S}{N} & = -\ln\det(\partial_\tau+\varepsilon_{k+A}-\mu+\Sigma)+\frac{1}{2}\ln\det(-\partial_\tau^2+\omega_q^2+m_b^2-\Pi) \\
       & -\int\rd^3 x\rd^3 x'\left(\Sigma(x',x)G(x,x')-\frac{1}{2}\Pi(x',x)D(x,x')\right) \\
       & +\int\rd^3 x\rd^3 x' \frac{g'^2}{2}G(x,x')G(x',x)D(x,x')\bar{\delta}(\vec{x}-\vec{x}')\,.
  \end{split}
\end{equation} Here $G$ is the fermion Green's function, $\Sigma$ is the fermion self-energy, $D$ is the boson Green's function and $\Pi$ is the boson self-energy. The determinant is acting on the functionals of spacetime functions. Here and later we use $x$  to denote spacetime coordinates, and use $\vec{x}$ to denote its spatial part. Also we will use $\bar{\delta}$ to represent spatial delta functions.

\FloatBarrier
\subsection{Saddle point equations}

  Taking the functional derivatives of Eq.\eqref{eq:GSprime}, we obtain the saddle point equations
  \begin{eqnarray}
    G(x,y) &=& (-\partial_\tau+\mu-\varepsilon_{k+A}-\Sigma)^{-1}(x,y)\,, \label{eq:SDp1} \\
       D(x,y)  &=& (-\partial_\tau^2+\omega_q^2-\Pi)^{-1}(x,y)\,,\label{eq:SDp2} \\
       \Sigma(x,y)  &=& g'^2 G(x,y)D(y,x)\bar{\delta}(\vec{x}-\vec{y})\,,\label{eq:SDp3} \\
         \Pi(x,y)  &=& -g'^2 G(x,y)G(y,x)\bar{\delta}(\vec{x}-\vec{y}) \,,\label{eq:SDp4}
  \end{eqnarray}where the inverse is in the functional sense.

  \subsubsection{Landau level basis}

  We proceed to analyze these equations in the presence of magnetic field. Unless otherwise stated we will work at zero temperature. It's useful to transform to the Landau level basis, where Eq.\eqref{eq:SDp1} will be come diagonal. Since $\Sigma(x,y)$ is proportional to spatial delta functions, together with (average) translational invariance, it is proportional to the identity matrix in the Landau level basis, i.e. it is only a function of frequency \cite{Aldape2022}. We can now expand the Green's function as
  \begin{equation}\label{eq:G=Gnphiphi}
    G(\vec{x},\vec{y},i\omega)=\sum_{n}\int\frac{\rd k}{2\pi} G_n(i\omega)\phi_{nk}(\vec{x})\phi_{nk}^*(\vec{y})\,,
  \end{equation}
   \begin{equation}\label{eq:GLandau}
     G_n(i\omega)=\frac{1}{i\omega+\mu-\left(n+\frac{1}{2}\right)\omega_c-\Sigma(i\omega)}\,.
   \end{equation}
   Here we have Fourier transformed the imaginary time $\tau$ into Matsubara frequency $i\omega$. $\omega_c=eB/m$ is the bare cyclotron frequency and $m$ is the bare band mass.  The sum over $n$ is over Landau levels and the integral over $k$ is over the residual momentum in the Landau gauge. The functions $\phi_{nk}$ are the Landau level wavefunctions:
  \begin{equation}\label{}
    \phi_{nk}(\vec{x})=\frac{e^{ikx_1}}{\sqrt{\ell_B}} \varphi_n\left(\frac{x_2}{\ell_B}-k\ell_B\right)\,,
  \end{equation} where $\ell_B=1/\sqrt{eB}$ and $\varphi_n(x)$ is the normalized Hermite function
  \begin{equation}\label{}
    \varphi_n(z)=\frac{1}{\sqrt{2^nn!\sqrt{\pi}}}H_n(z) e^{-\frac{z^2}{2}}\,,
  \end{equation}  where $H_n(x)$ is the physicist's Hermite polynomial. Here $x_1$, $x_2$ denote the two components of spatial coordinate $\vec{x}$.

  The $k$-integral in \eqref{eq:G=Gnphiphi} can be calculated by mapping to simple harmonic oscillator  to yield
  \begin{equation}\label{eq:Gxy}
    G(\vec{x},\vec{y},i\omega)=\sum_{n} G_n(i\omega) \frac{e^{i\theta_B(\vec{x},\vec{y})}}{2\pi\ell_B^2} \exp\left(-\frac{|\vec{x}-\vec{y}|^2}{4\ell_B^2}\right)L_n\left(\frac{|\vec{x}-\vec{y}|^2}{2\ell_B^2}\right)\,.
  \end{equation} Here $\theta_B(\vec{x},\vec{y})=\frac{(x_1-y_1)(x_2+y_2)}{2\ell_B^2}$ is a gauge-dependent phase factor that breaks translational invariance because $G$ is not gauge invariant, and $L_n$ is the Laguerre polynomial. A similar result has appeared before in Ref.~\cite{Barci2018}.

  \subsubsection{Boson self-energy}

  We now simplify Eq.\eqref{eq:SDp4}; due to the delta function we can set $\vec{x}=\vec{y}$ in Eq.\eqref{eq:Gxy}, and consider an auxiliary Green's function
  \begin{equation}\label{}
    \bar{G}(i\omega)=-\frac{\omega_c}{\pi} \sum_{n}G_n(i\omega)\,.
  \end{equation} In the large $k_F$ limit, most contributions of the sum arise from near the Fermi surface, and we can approximate the sum to be over all integers, yielding
  \begin{equation}\label{eq:bGgp}
    \bar{G}(i\omega)=\tan\frac{i\omega+\mu-\Sigma(i\omega)}{\omega_c}\,.
  \end{equation}

  The boson self-energy then reads
  \begin{equation}\label{eq:Pi=bGbG}
    \Pi(i\Omega)=-\frac{g'^2m^2}{4} \int\frac{\rd \omega}{2\pi} \bar{G}(i\omega)\bar{G}(i\omega+i\Omega)\,.
  \end{equation}

  These equations are consistent with the zero field limit. In the $\omega_c\to 0$ limit, $\bar{G}(i\omega)\to i\sgn\omega$, and we can compute the integral \eqref{eq:Pi=bGbG} and obtain
  \begin{equation}\label{}
    \Pi_{\omega_c\to 0}(i\Omega)-\Pi_{\omega_c\to 0}(0)=-\frac{g'^2m^2}{4\pi}|\Omega|\,.
  \end{equation}

  \subsubsection{Fermion self-energy}

  Next we simplify Eq.\eqref{eq:SDp3}. We will need the local boson Green's function, which is
  \begin{equation}\label{}
    \bar{D}(i\nu)=\int \frac{\rd^2 \vec{q}}{(2\pi)^2} \frac{1}{\nu^2+q^2-\bar{\Pi}(i\nu)}=\frac{1}{4\pi}\ln\frac{\nu^2-\bar{\Pi}(i\nu)+\Lambda_q^2}{\nu^2-\bar{\Pi}(i\nu)}\,.
  \end{equation} Here the boson is critical by setting $\bar{\Pi}(i\nu)=\Pi(i\nu)-\Pi(0)$ through tuning the boson mass, and $\Lambda_q$ is the momentum cutoff for the boson.

  The fermion Green's function then reads
  \begin{equation}\label{eq:Sigma=bDbG}
    \Sigma(i\omega)=-\frac{g'^2m}{2}\int\frac{\rd \nu}{2\pi}\bar{D}(i\nu)\bar{G}(i\nu+i\omega)\,.
  \end{equation}

  In the $\omega_c\to 0$ limit, we recover the marginal Fermi liquid result
  \begin{equation}\label{}
      \Sigma_{\omega_c\to 0}(i\omega)=-i\frac{g'^2 m\omega}{8\pi^2}\ln\frac{e \Lambda_q^2}{\gamma|\omega|}\,,
    \end{equation} agreeing with \cite{Patel2023}.
\FloatBarrier
  \subsection{Numerical solution of the saddle point equations}

  In this part, we analyze the saddle point equations \eqref{eq:SDp1}-\eqref{eq:SDp4} numerically in the real frequency domain.

  \subsubsection{Method}

   We adapt the saddle point equations with additional cutoffs to allow for numerical solution. For the auxiliary Green's function $\bar{G}$, we insert bandwidth cutoffs for the Landau level sum:
  \begin{equation}\label{eq:SDpbarG}
    \begin{split}
      &\bar{G}(i\omega)=-\frac{\omega_c}{\pi}\sum_{n=n_-}^{n_+-1}\frac{1}{i\omega+\mu-\Sigma(i\omega)-(n+1/2)\omega_c}\\
      &=-\frac{1}{\pi}
      \left[\psi\left(\frac{1}{2}+n_--\frac{i\omega-\Sigma(i\omega)}{\omega_c}\right)-\psi\left(\frac{1}{2}+n_-+\frac{i\omega-\Sigma(i\omega)}{\omega_c}\right)\right]
    \,.\end{split}
    \end{equation}Here the cutoffs are set by $n_+=-n_-=W/(2\omega_c$) where $W$ is the bandwidth, and we focus on half filling by setting $\mu=0$.

    Next, we analytically continue to real frequencies, yielding the retarded Green's function
    \begin{equation}
    \label{eq:SDpn1}
      \bar{G}_R(\omega)=-\frac{1}{\pi}
      \left[\psi\left(\frac{1}{2}+n_--\frac{\omega+i\eta-\Sigma_R(\omega)}{\omega_c}\right)-\psi\left(\frac{1}{2}+n_-+\frac{\omega+i\eta-\Sigma_R(\omega)}{\omega_c}\right)\right]\,.
    \end{equation} Here $\eta$ is theoretically a positive infinitesimal quantity, but in numerics it will be a small finite number.

    Similarly, the local boson's Green's function can be written as
    \begin{equation}\label{eq:SDpn2}
      \bar{D}_R(\nu)=\frac{1}{4\pi}\ln \frac{(\eta-i\nu)^2-\bar{\Pi}_R(\nu)+\Lambda_q^2}{(\eta-i\nu)^2-\bar{\Pi}_R(\nu)}\,.
    \end{equation}

    The self energies are more conveniently expressed in the time-domain (the Fourier convention is $A(t)=\int\frac{\rd \omega}{2\pi} A(\omega)e^{-i\omega t}$). Following \cite{Schmalian1996}, we compute the Matsubara summation in Eqs.\eqref{eq:Pi=bGbG} and \eqref{eq:Sigma=bDbG} and deform the integration contour to the real axis, obtaining
    \begin{equation}\label{eq:SDpn3}
    \Pi_R(t)=-\frac{g'^2m^2}{2}\Re\left[\bar{G}_R(t)\tilde{G}(t)^*\right]\,,
  \end{equation}
  \begin{equation}\label{eq:SDpn4}
    \Sigma_R(t)=\frac{g'^2m}{2}\left[\bar{G}_R(t)\tilde{D}(t)^*-\tilde{G}(t)\bar{D}_R(t)\right]\,.
  \end{equation} Here we have introduced two additional functions $\tilde{G}$ and $\tilde{D}$, which are related to the spectral functions of $\bar{G}$ and $\bar{D}$
  \begin{eqnarray}
    \tilde{G}(\omega) &=& -2n_F(\omega)\Im \bar{G}_R(\omega)\,, \\
    \tilde{D}(\omega) &=& -2n_B(\omega)\Im \bar{D}_R(\omega)\,,
  \end{eqnarray} and $n_F(n_B)$ is the Fermi-Dirac (Bose-Einstein) function.

  Eqs.\eqref{eq:SDpn1}, \eqref{eq:SDpn2}, \eqref{eq:SDpn3}, and \eqref{eq:SDpn4} can be solved numerically using the standard fast Fourier transform + iteration method. The frequency domain is represented by a grid $[-\omega_\text{max},\omega_\text{max})$ with spacing $\Delta\omega$. The time domain is discretized accordingly to $[0,T_\text{max})$ with spacing $\delta t$. The time grid is related to frequency grid via the relations $T_\text{max}={2\pi}/({\Delta \omega})$ and $\Delta t={\pi}/{\omega_\text{max}}$.

  \subsubsection{Result}\label{sec:gpnumres}

\begin{figure}
    \centering
    \includegraphics[width=\textwidth]{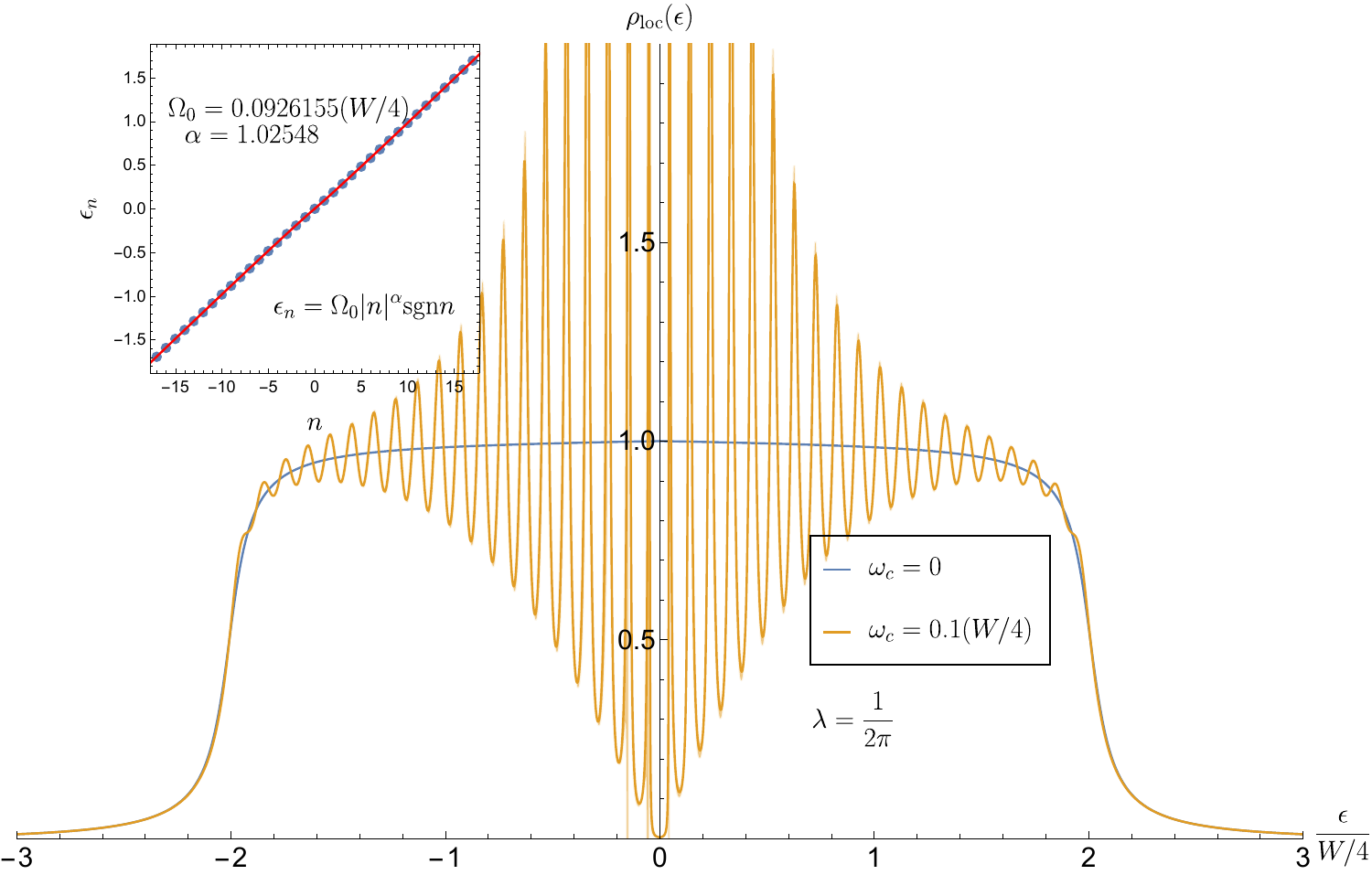}
    \caption{The fermion local density of states $\rho_\text{loc}(\epsilon)$ at small coupling $\lambda=1/(2\pi)$. The blue curve shows the zero-field result and the yellow curve shows the finite-field result with bare cyclotron frequency $\omega_c=0.1$. The inset plots the position of minima $\epsilon_n$ versus index $n$ together with a power-law fitting. The extracted power-law is close to linear and minima spacing $\Omega_0$ is also close to bare cyclotron mass $\omega_c$. Because of the small coupling, a decade of oscillations in $\rho_\text{loc}(\epsilon)$ is visible.}\label{fig:gprime_Gloc1}
  \end{figure}

\begin{figure}
    \centering
    \includegraphics[width=\textwidth]{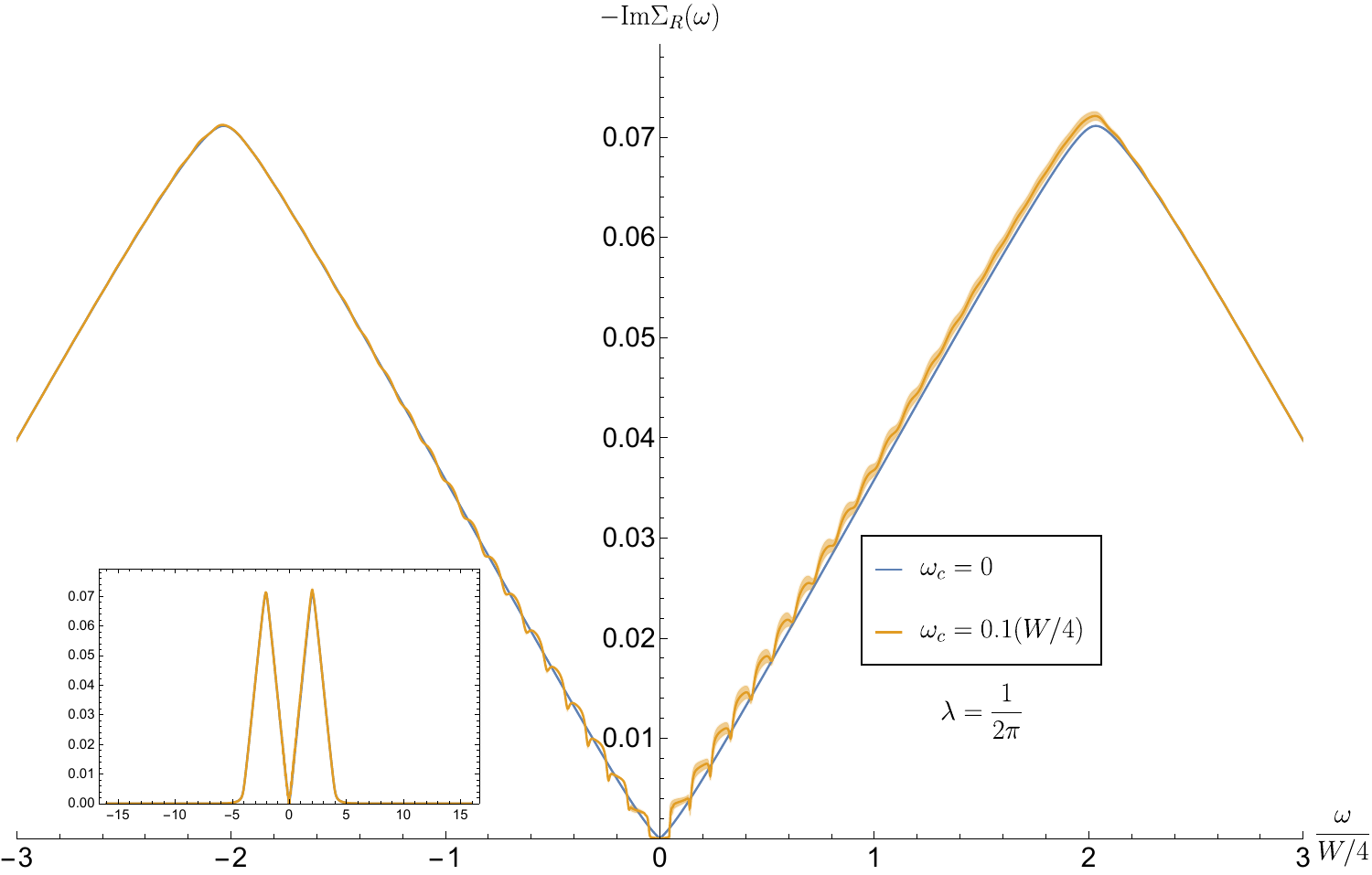}
    \caption{The imaginary part of fermion self-energy $\Im\Sigma_R(\omega)$ at small coupling $\lambda=1/(2\pi)$. The blue curve shows the zero-field result which asymptotes to $|\omega|$ at small frequencies, and the yellow curve shows the finite-field result with bare cyclotron frequency $\omega_c=0.1$. The shaded regions represent the numerical uncertainty of the finite-field result. The inset plots the results on a larger frequency scale and confirms that $\Im\Sigma_R$ vanishes at frequencies above the bandwidth. }\label{fig:gprime_Sigma1}
  \end{figure}

\begin{figure}
    \centering
    \includegraphics[width=\textwidth]{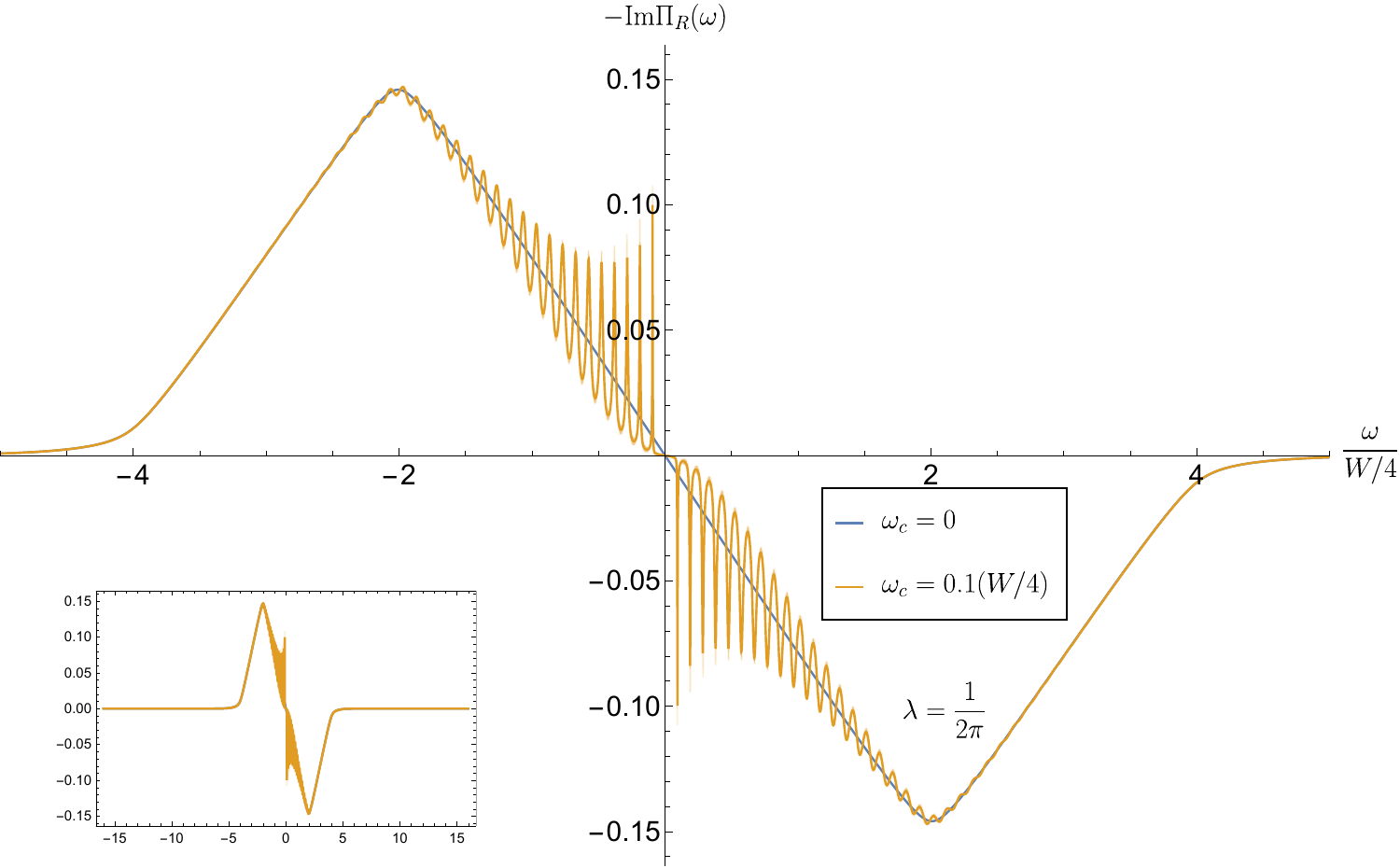}
    \caption{The imaginary part of the boson self-energy $\Im\Pi_R(\omega)$ at small coupling $\lambda=1/(2\pi)$. The blue curve shows the zero-field result which is linear in $\omega$ and the yellow curve shows the finite-field result with bare cyclotron frequency $\omega_c=0.1$. The shaded regions represent the numerical uncertainty of the finite-field result. The inset plots the result on a larger frequency scale. }\label{fig:gprime_Pi1}
  \end{figure}

\begin{figure}
    \centering
    \includegraphics[width=\textwidth]{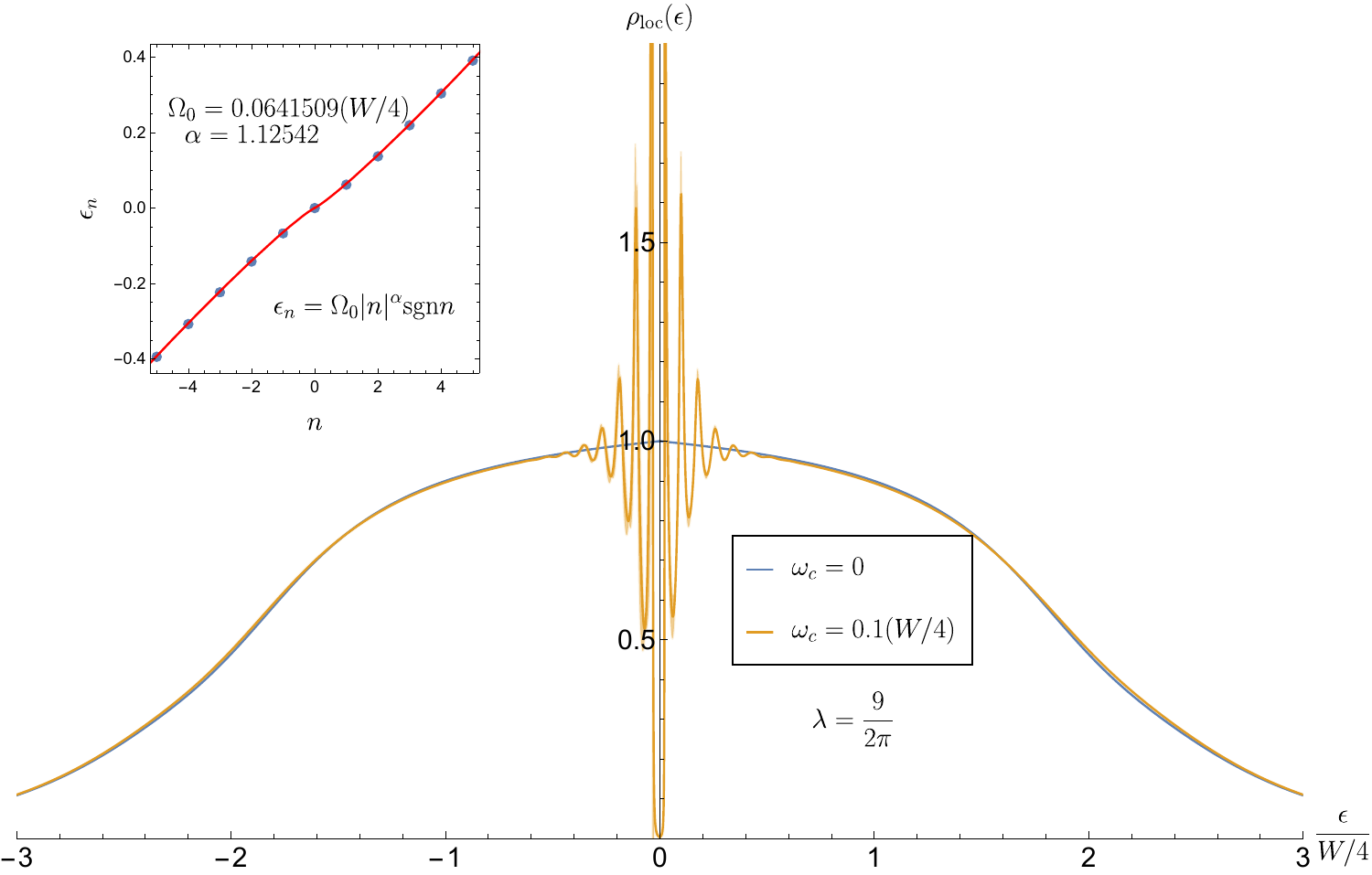}
    \caption{Same as Fig.~\ref{fig:gprime_Gloc1} but at a larger coupling $\lambda=9/(2\pi)$. At this stronger coupling, the oscillations in $\rho_\text{loc}(\epsilon)$ as a function of $\epsilon$ has been damped with less minima visible. As the inset shows, the positions of oscillation minima $\epsilon_n$ has deviated from the linear distribution and the typical spacing $\Omega_0$ is significantly smaller than the bare cyclotron frequency $\omega_c$.}\label{fig:gprime_Gloc2}
  \end{figure}

\begin{figure}
    \centering
    \includegraphics[width=\textwidth]{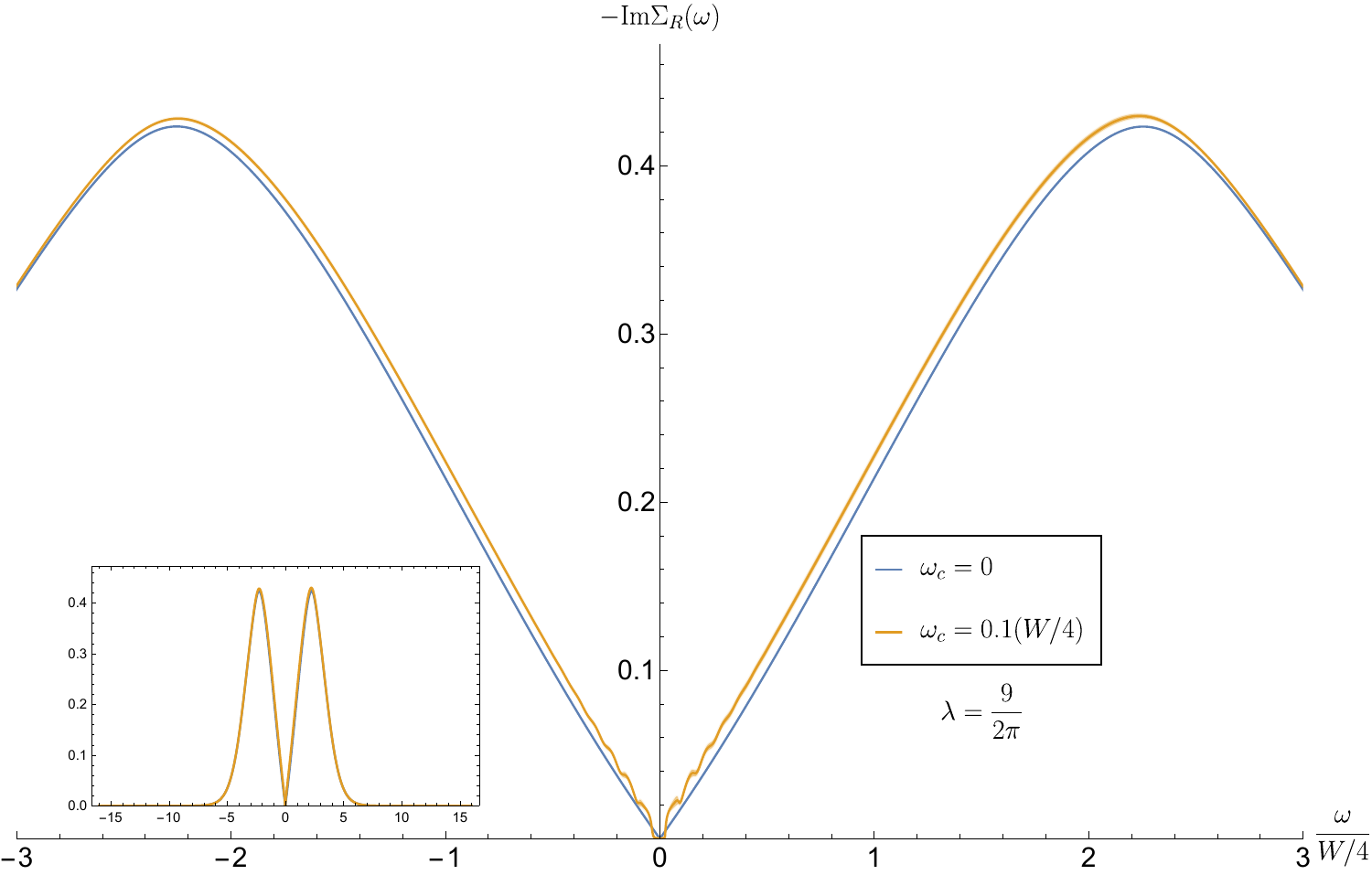}
    \caption{Same as Fig.~\ref{fig:gprime_Sigma1} but at a larger coupling $\lambda=9/(2\pi)$.}\label{fig:gprime_Sigma2}
  \end{figure}

\begin{figure}
    \centering
    \includegraphics[width=\textwidth]{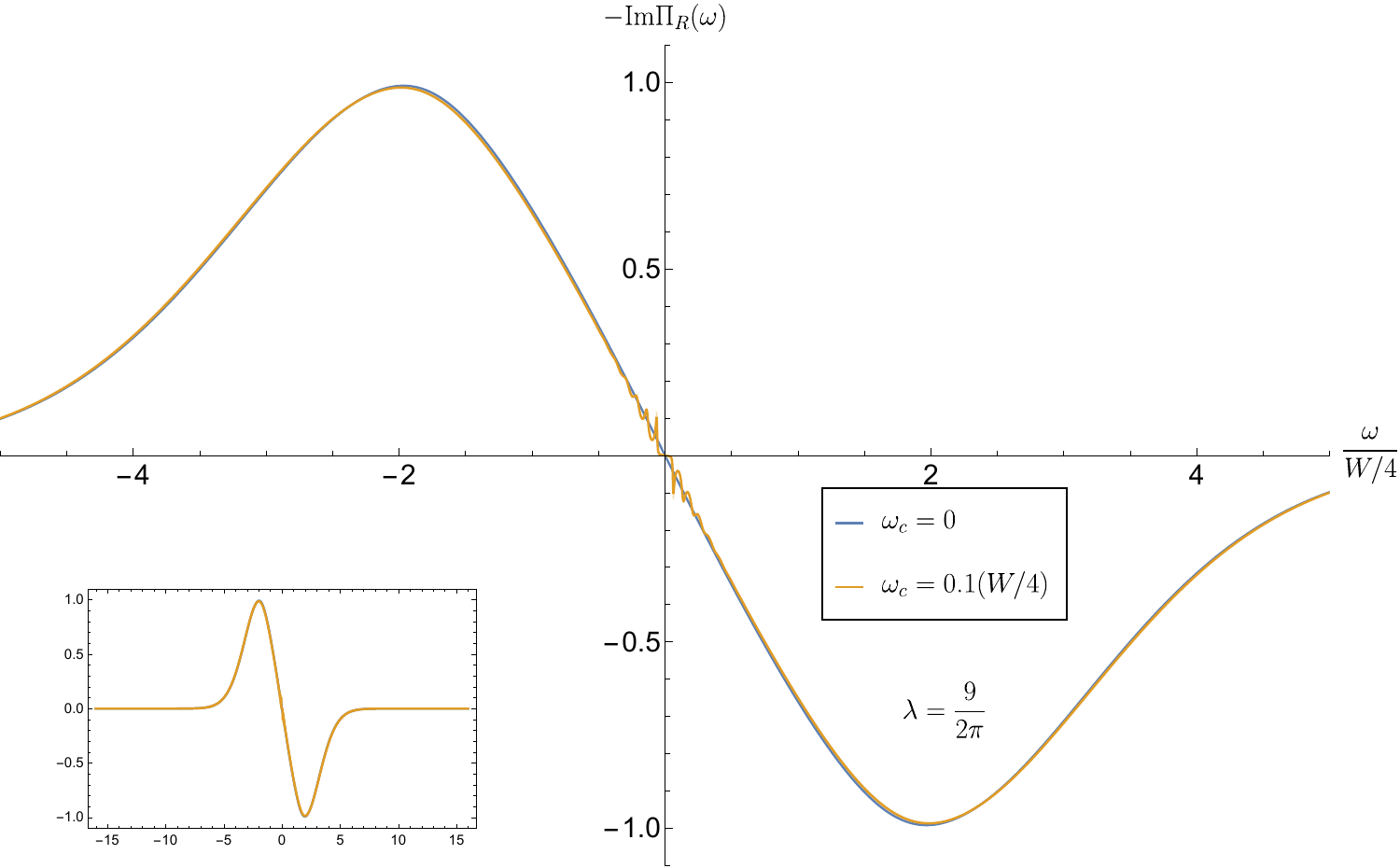}
    \caption{Same as Fig.~\ref{fig:gprime_Pi1} but at a larger coupling $\lambda=9/(2\pi)$. }\label{fig:gprime_Pi2}
  \end{figure}

    In this section we present the results of the numerical solution. The parameters we used are $\omega_\text{max}=16$, $\Delta\omega=0.001$, $W=4$, $\Lambda_q=2$, $\eta=10^{-4}$, $\omega_c=0.1$, $m=1$. We present the result in terms of the fermion local density of states $\rho_\text{loc}(\omega)=\Im \bar{G}_R(\omega)$ (Figs.~\ref{fig:gprime_Gloc1} and \ref{fig:gprime_Gloc2}), the imaginary part of the fermion self-energy $\Im\Sigma_R(\omega)$ (Figs.~\ref{fig:gprime_Sigma1} and \ref{fig:gprime_Sigma2}) and the imaginary part of the boson self-energy $\Im \Pi_R(\omega)$ (Figs.~\ref{fig:gprime_Pi1} and \ref{fig:gprime_Pi2}). We have converted the disordered Yukawa-SYK coupling $g'$ to a dimensionless parameter $\lambda=mg'^2/(2\pi)$. We verified the fermionic sum-rule that $\left|1-\frac{1}{W}\int \rd\omega \rho_\text{loc}(\omega)\right|<10^{-3}$. We also verified that in the zero-field case $\omega_c=0$, we recover the marginal Fermi liquid $|\omega|$ behavior in $\Im\Sigma_R(\omega)$ at low energies.

    In Figs. \ref{fig:gprime_Gloc1} through \ref{fig:gprime_Pi1}, we used a small coupling constant $g'=1$, and in Fig.~\ref{fig:gprime_Gloc2} through Fig.~\ref{fig:gprime_Pi2} we chose a larger coupling $g'=3$. In the former case $|\Im\Sigma_R(\omega)|\ll|\omega|$ and the system is similar to free fermions, and in the latter case the marginal fermi liquid self-energy is more pronounced. We found that the marginal fermi liquid and oscillatory behaviors are in competence with each other that oscillatory effects are strong only when the marginal fermi liquid is weak. This behavior can be qualitatively understood by rewriting Eq.\eqref{eq:bGgp} as
    \begin{equation}\label{eq:bGgp2}
      \bar{G}_R(\omega)=\frac{\sin\left(\frac{\omega+\mu-\Re \Sigma_R}{\omega_c/2}\right)+i\sinh\frac{2\Im \Sigma_R}{\omega_c}}{\cos\left(\frac{\omega+\mu-\Re \Sigma_R}{\omega_c/2}\right)+\cosh\frac{2\Im \Sigma_R}{\omega_c}}\,.
    \end{equation} According to Eq.\eqref{eq:bGgp2} the oscillatory part of $\bar{G}_R(\omega)$ is controlled exponentially by $\Im\Sigma_R/\omega_c$.  When $|\Im\Sigma_R|>\omega_c$, it is justified to do a first order expansion in $\exp(-|\Im\Sigma_R/\omega_c|)$, by only retaining oscillations in the Green's function $\bar{G}_R(\omega)$ and using the zero-field limit for $\Sigma_R(\omega)$.



\FloatBarrier
  \subsection{Local density of states}\label{sec:gpldos}

    In this part, we analyze the local density of states of the $g'$-model. Following \cite{Chowdhury2018}, we compute the local Green's function
        \begin{equation}\label{}
          G_\text{loc}(i\omega)=\frac{1}{2\pi\ell_B^2}\sum_{n=0}^{\infty}\frac{1}{i\omega+\mu-\Sigma(i\omega)-\left(n+\frac{1 }{2}\right)\omega_c}
        \end{equation}
        We assume the large Fermi surface limit such that the Landau level sum can be extended to $-\infty$. Using the Poisson resummation formula
        we obtain
        \begin{equation}\label{}
          G_\text{loc}(i\omega)=\frac{eB}{2\pi}\int_{-\infty}^{\infty} \rd n \sum_{k=-\infty}^{\infty} \frac{e^{2\pi i kn}}{i\omega+\mu-\Sigma(i\omega)-\left(n+\frac{1}{2}\right)\omega_c}\,.
        \end{equation}
         We analytically continue to the retarded Green's function $i\omega\to \omega+i\eta$, and evaluate the $n$-integral using contour methods, we obtain the local density of states
        \begin{equation}\label{eq:Alocoscillation}
          \rho_\text{loc}(\omega)=-2\Im G_{R,\text{loc}}(\omega)=m+2m \sum_{k=1}^{\infty} e^{-\frac{2\pi k|\Sigma_R''(\omega)|}{\omega_c}}\cos\left(\frac{2\pi k}{\omega_c}(\omega+\mu-\Sigma_R'(\omega))\right)\,.
        \end{equation}
         Here we have decomposed the self-energy into real and imaginary parts with $\Sigma_R(\omega)=\Sigma_R'(\omega)+i\Sigma_R''(\omega)$.

        The experimental implications of Eq.\eqref{eq:Alocoscillation} are the following:
        \begin{itemize}
          \item  Because $\mu\gg |\omega-\Sigma_R(\omega)|$, the quantum oscillation in $1/B$ will still show a frequency determined from the Fermi surface area, similar to a conventional Fermi liquid.
          \item However, in the marginal Fermi liquid or non-Fermi liquid regime $|\Sigma_R(\omega)|\gg |\omega|$, so if a scanning tunneling microscopy experiment is performed, it can be observed that the energy ($\omega$) dependence is no longer quasi-periodic (due to $\Re\Sigma_R$) with period $\omega_c$, and less oscillations in $\omega$ can be visible (due to $\Im\Sigma_R$), as shown in Figs.~\ref{fig:gprime_Gloc1} and \ref{fig:gprime_Gloc2}.

          \item Additionally, the Dingle factor $\exp\left(-\frac{2\pi k|\Sigma_R''(\omega)|}{\omega_c}\right)$ can also be sensitive to NFL/MFL physics. In particular, the NFL/MFL interaction will contribute a $\omega$-dependent Dingle factor, which is distinct from the $\omega$-independent contribution from potential disorder. It is also possible to differentiate the NFL/MFL contribution to Dingle factor from the FL contributions because the former should be much larger than the bare frequency $\omega$, and $\omega$ should be larger than the latter.

        \end{itemize}

\begin{figure}
    \centering
    \includegraphics{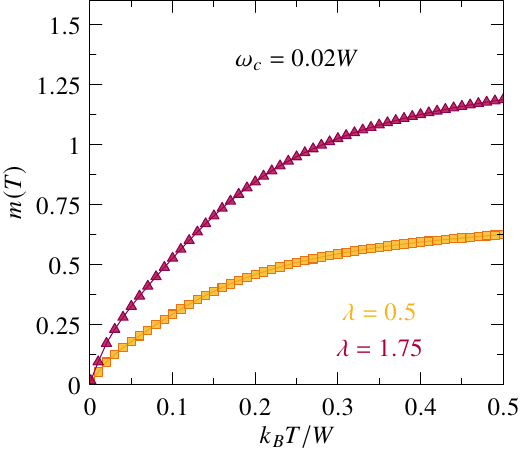}
    \caption{The bosonic thermal mass $m_b(T)$ at weak $(\lambda=0.5)$ and strong couplings $(\lambda=1.75)$. For both couplings, the bare boson mass $m_b$ is tuned to criticality in the limit $T\rightarrow 0^+$.}
    \label{fig:mT}
\end{figure}

\begin{figure}[htb ]
   \centering
   \includegraphics[width=0.85\textwidth]{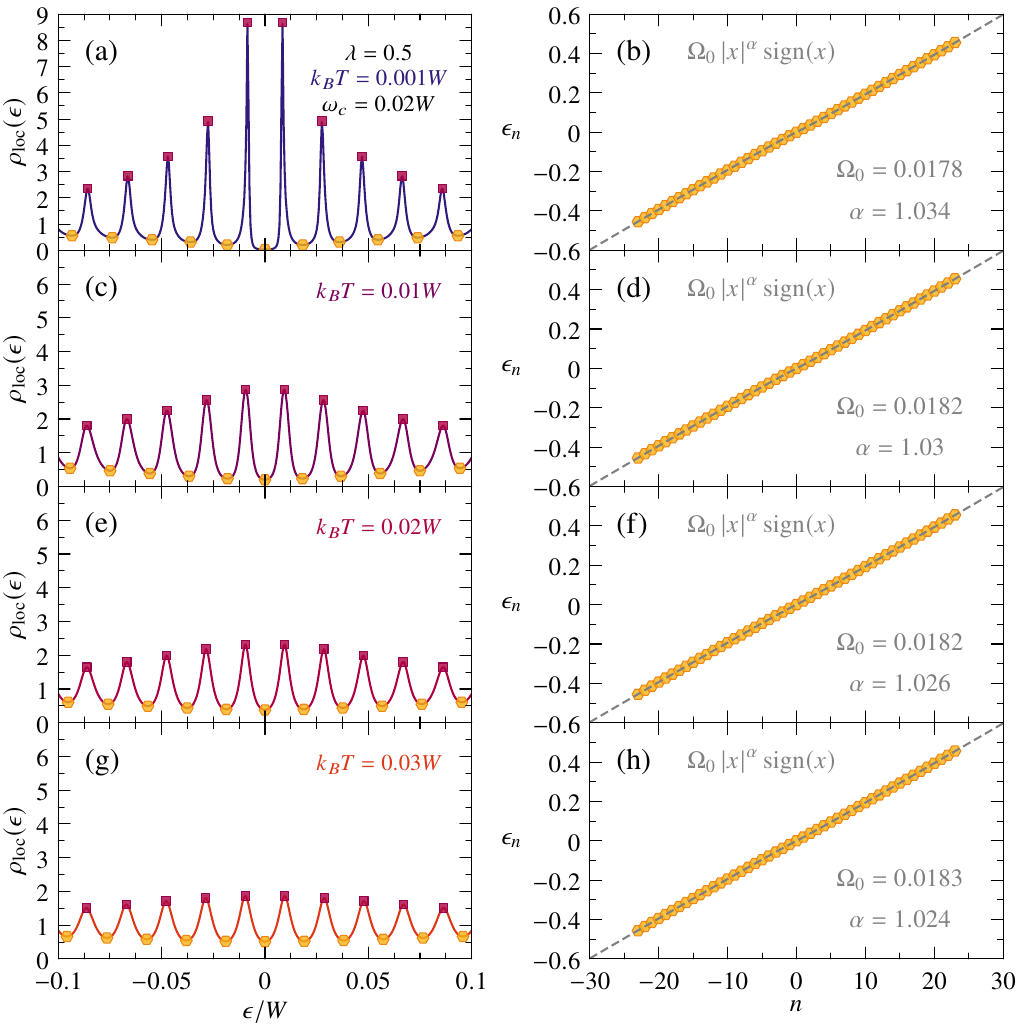}
   \caption{Left panels: Fermionic density of states at low energies for different temperature at $\omega_{c}=0.02W$
and dimensionless coupling constant $\lambda=0.5$, i.e. weak to moderate coupling. Right panels: location of the Landau-level peaks in the spectrum. Landau levels are essentially equal spaced.}
\label{fig:Fig_DOS_extremes_lambda0p5}
 \end{figure}

\begin{figure}[htb ]
   \centering
   \includegraphics[width=0.85\textwidth]{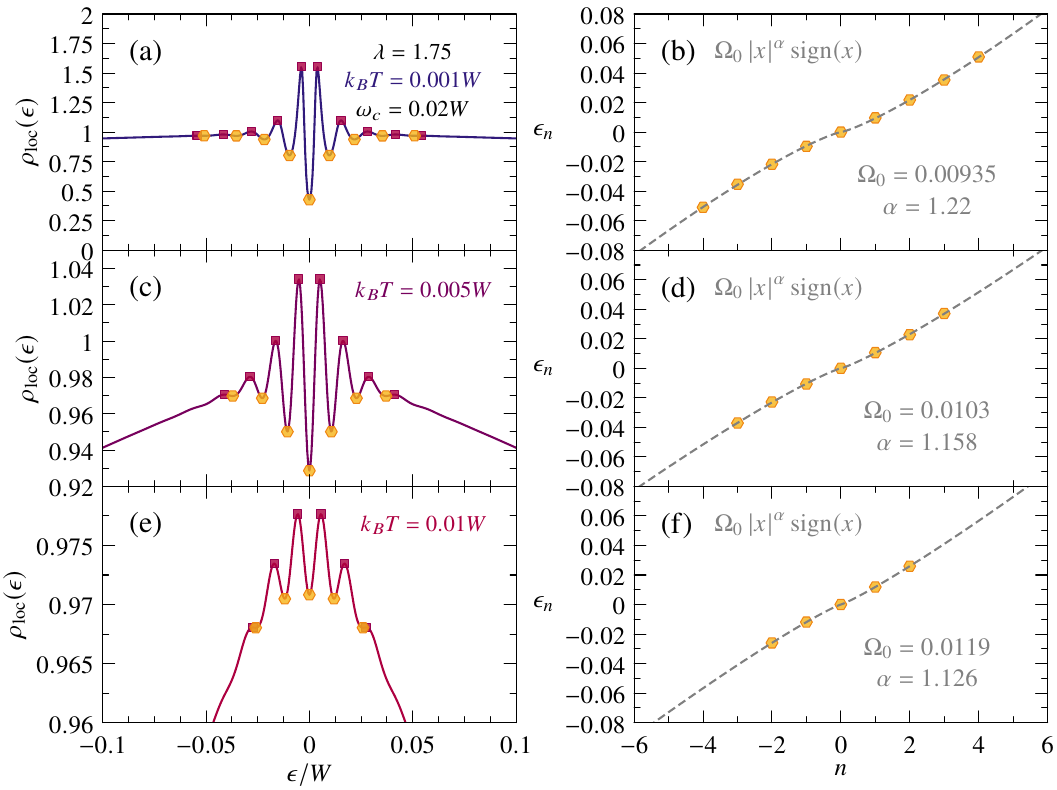}
   \caption{Left panels: Fermionic density of states at low energies for different temperature at $\omega_{c}=0.02W$
and dimensionless coupling constant $\lambda=1.75$, i.e. strong coupling. Right panels: location of the Landau-level peaks in the spectrum. Landau levels are  no-longer essentially equal spaced, an effect caused by the strong renormalization of the real part of the self energy as function of frequency. }
\label{fig:Fig_DOS_extremes_lambda1p75}
 \end{figure}

\joerg{Finally we analyze the single particle spectrum as a function
of temperature at the quantum critical point. Here we included a finite fermionic cut off energy
$W$ in the numerical analysis and performed the calculation for $\omega_{c}/W=0.02$ and two values of the dimensionless
coupling strength $\lambda=\frac{1}{2\pi}mg'^{2}$  at  varying $T/W$.  As shown in Figs.\ref{fig:Fig_DOS_extremes_lambda0p5}  and
 \ref{fig:Fig_DOS_extremes_lambda1p75}, we use $\lambda=0.5$, which
corresponds to weak to moderate  coupling, while $\lambda=1.75$ describes stronger coupling.  We observe the thermal suppression of the Landau-levels, in agreement with the $T$-dependent Dingle factor of Eq.~\eqref{eq:Alocoscillation}. The effect is again much more pronounced at strong coupling. Notice, even at weak coupling the oscillations drift very slightly with temperature, an effect that is absent if we ignore the $T$-dependence of the bosonic mass $m_b(T)$ (shown in Fig.~\ref{fig:mT}). The latter is obtained from the solution of the coupled large-$N$ equations as is a result of the $T$-dependence of the bosonic self energy $\Pi(0)$, that follows for the finite-$T$ version of Eq.~\eqref{eq:Pi=bGbG}.
}

\FloatBarrier
  \subsection{Magnetization}\label{sec:gpmag}
The magnetization can be calculated from the free energy via
        \begin{equation}\label{}
          M(B)=\frac{T}{V}\frac{\partial S}{\partial B}\,,
        \end{equation} where $S$ is the saddle-point action defined in \eqref{eq:GSprime}. At the saddle-point, all implicit dependence of $S$ on $G,\Sigma,D,\Pi$ are cancelled out and we only need to include the explicit $B$ dependence, i.e. the kinetic term.

        The magnetization formula is therefore \footnote{There is an additional contribution due to variation of the Landau level degeneracy with the magnetic field. However, this term is expected to be less significant in the low-field limit.}
        \begin{equation}\label{}
          M(B)=\frac{1}{2\pi\ell_B^2}\sum_n T\sum_{i\omega} G_n(i\omega)\frac{\partial \varepsilon_n}{\partial B}\,,
        \end{equation}  where $\varepsilon_n=(n+1/2)\omega_c$. We rewrite the summation using spectral functions to obtain
        \begin{equation}\label{}
          M(B)=\frac{e^2 B}{2\pi m} \sum_n \left(n+\frac{1 }{2}\right) \int \frac{\rd z}{2\pi}n_F(z)\rho_n(z)\,,
        \end{equation} where the spectral function $\rho_n(z)$ is
        \begin{equation}\label{}
          \rho_n(z)=-2\Im G_{R,n}(z)=\frac{-2\Sigma_R''(z)}{\left(z+\mu-\Sigma_R'(z)-(n+1/2)\omega_c\right)^2+\Sigma_R''(z)^2}\,.
        \end{equation} We apply the Poisson resummation formula to $n$ and obtain
        \begin{equation}\label{}
          M(B)=\frac{m\mu}{2\pi B} \sum_{k=-\infty}^{\infty} \int \rd z n_F(z) e^{\frac{2\pi i k}{\omega_c}(z+\mu-\Sigma'_R(z))} e^{-\frac{2\pi}{\omega_c}|k \Sigma''_R(z)|}\left[1+\frac{z-\Sigma'_R(z)+i|\Sigma_R''(z)|}{\mu}\right]\,.
        \end{equation} Since $\mu\gg |z-\Sigma_R(z)|$, we expect the magnetization oscillation in $1/B$ to be similar to a FL.

\FloatBarrier
    \subsection{Transport}\label{sec:gptransport}
    \subsubsection{Formula for conductivity}

  We start from the Kubo formula
  \begin{equation}\label{}
    \sigma^{\mu\nu}(\omega)=\frac{1}{S}\left.\frac{\Pi_{jj}^{\mu\nu}(i\Omega)-\Pi_{jj}^{\mu\nu}(0)}{\Omega}\right\vert_{i\Omega\to \omega+i0}\,.
  \end{equation} Here $S$ is the area of the system, and  the current-current correlator $\Pi_{jj}^{\mu\nu}$ has the same form as the previous paper \cite{Guo2022}:
  \begin{equation}\label{}
    \Pi_{jj}^{\mu\nu}=\Gamma^{\mu,T}\frac{1}{W_{\Sigma}^{-1}-W_\text{MT}-W_\text{AL}} \Gamma^{\nu}\,.
  \end{equation} Here $\Gamma^\mu$ is the bare current vertex which we describe below. $W_\Sigma, W_\text{MT}$ and $W_\text{AL}$ are functional kernels that act on $\Gamma^\mu$. As discussed in the previous work \cite{Guo2022}, $W_\Sigma$ corresponds to attaching a pair of fermion Green's functions, and $W_\text{MT}$, $W_\text{AL}$ generate the Maki-Thompson and Aslamazov-Larkin diagrams respectively.

  The bare velocity vertices in first quantized picture are
  \begin{equation}\label{}
    v_x=m^{-1}\left(-i\partial_x-eBy\right)=-\frac{1}{m\ell_B}\frac{a+a^\dagger}{\sqrt{2}}\,,
  \end{equation}
  \begin{equation}\label{}
    v_y=-m^{-1}i\partial_y=\frac{1}{m\ell_B}\frac{i(a^\dagger-a)}{\sqrt{2}}\,.
  \end{equation} Here $a,a^\dagger$ are ladder operators acting on the Landau levels $\phi_{nk}$.

  Therefore, the bare current vertices are given by
  \begin{equation}\label{}
    \Gamma^x_{n'k',nk}=-\frac{1}{\sqrt{2}m\ell_B}\left(\sqrt{n}\delta_{n,n'+1}+\sqrt{n'}\delta_{n',n+1}\right)2\pi\delta(k-k')\,,
  \end{equation}
  \begin{equation}\label{}
    \Gamma^y_{n'k',nk}=\frac{i}{\sqrt{2}m\ell_B}\left(\sqrt{n'}\delta_{n',n+1}-\sqrt{n}\delta_{n,n'+1}\right)2\pi\delta(k-k')\,.
  \end{equation} Here $n'(n)$ denote the Landau level index of the outgoing(incoming) fermion.

  We take the linear combination $\Gamma^\pm=(\Gamma^x\pm i \Gamma^y)/2$ and consider
  \begin{equation}\label{eq:bareGammap}
    \Gamma^+_{n'k',nk}=-\frac{1}{\sqrt{2}m\ell_B}\sqrt{n'}\delta_{n',n+1}2\pi \delta(k-k')\,,
  \end{equation}
  \begin{equation}\label{eq:bareGammam}
    \Gamma^-_{n'k',nk}=-\frac{1}{\sqrt{2}m\ell_B}\sqrt{n}\delta_{n,n'+1}2\pi \delta(k-k')\,.
  \end{equation}
  Conductivity in the above basis $\sigma^{-+}$ and $\sigma^{+-}$ correspond to the right/left circular polarizations respectively.

  The kernel $W_\Sigma$ is diagonal in the space of two-point functions:
  \begin{equation}\label{}
    W_\Sigma[F](x',y')=G(x',x)F(x,y)G(y,y')\,,
  \end{equation} or in the Landau level-frequency space it reads
  \begin{equation}\label{}
    W_\Sigma[F]_{n',n}(i\omega,i\Omega)=G_{n'}(i\omega+i\Omega/2) F_{n',n}(i\omega,i\Omega) G_n(i\omega-i\Omega/2)\,.
  \end{equation}

  We continue to evaluate $\sigma^{-+}(i\Omega)$ by treating the MT and the AL kernels.
   \subsubsection{MT diagram}

  The MT diagram kernel is
  \begin{equation}\label{}
    W_\text{MT}[F](x,y)=g'^2 D(x,y)F(x,y)\delta(\vec{x}-\vec{y})\,.
  \end{equation} Because of the $\delta$-function, the bare vertex is evaluated at coincidental spatial points. Using Eq.\eqref{eq:bareGammap} and transforming from Landau level basis to real space basis, we have
  \begin{equation}\label{}
    \Gamma^{+}(\vec{x},\vec{x})=-\frac{1}{\sqrt{2}m\ell_B}\sum_n \int \frac{\rd k}{2\pi}\sqrt{n+1}\frac{1}{\ell_B} \varphi_{n+1}(x_2/\ell_B-k\ell_B)\varphi_{n}(x_2/\ell_B-k\ell_B)=0\,,
  \end{equation} due to the orthogonality of the function $\varphi_n$. Therefore $W_\text{MT}[\Gamma^{+}]=0$. Similar reasonings apply to $\Gamma^{-}$ as well.

  \subsubsection{AL diagram}
   In real space, the AL kernel reads
  \begin{equation}\label{}
    W_\text{AL}[F](x,y)=-g^4\int\rd^3 x' \rd^3 y' G(x,y)D(x,x')D(y',y)\left[ F(x',y')G(y',x')+F(y',x') G(x',y')\right]\delta(\vec{x}'-\vec{y}')\delta(\vec{x}-\vec{y})\,,
  \end{equation} which, for the same reason as the MT diagram,  also vanishes when evaluated on the bare vertices $\Gamma^{\pm}$.

  \subsubsection{Conductivity}

  Therefore, the conductivity is given by a single fermion bubble. The polarization bubble is
  \begin{equation}\label{}
    \Pi_{jj}^{-+}(\Omega)=\frac{1}{2 m^2\ell_B^2} g_L\int \frac{\rd z_1\rd z_2}{(2\pi)^2}\frac{n_F(z_1)-n_F(z_2)}{z_1-z_2-\Omega-i0} \sum_n (n+1)A_{n+1}(z_1)A_n(z_2)\,,
  \end{equation} from which we obtain the DC conductivity
  \begin{equation}\label{eq:sigmaDC}
    \Re\sigma^{-+}_\text{DC}=\frac{e^2}{4m^2\ell_B^2 S}g_L \int \frac{\rd z}{2\pi}\left(-n_F'(z)\right)\sum_{n}(n+1)A_{n+1}(z)A_n(z)\,.
  \end{equation} Here $A_n(z)=-2\Im G_{Rn}(z)$ is the fermion spectral weight. $g_L=eBS/(2\pi)$ is the Landau level degeneracy which arises from the $k$-integral. The summation over $n$ can be evaluated, yielding
  \begin{equation}\label{eq:sigmaxx}
  \begin{split}
    \sigma^{xx}_\text{DC}=2\Re \sigma^{-+}_\text{DC}=\frac{e^2}{4\pi}\int\frac{\rd z}{2\pi} \frac{\left(-4n_F'(z)\right)\Sigma''(z)}{4\left[\Sigma''(z)\right]^2+\omega_c^2}\left[2 \Sigma''(z)+2(z+\mu-\Sigma'(z))\Im \psi\left(\frac{1}{2}-\frac{z-\Sigma'(z)+\mu}{\omega_c}+\frac{i \Sigma''(z)}{\omega_c}\right)\right]\,,
  \end{split}
  \end{equation} where $\Sigma_R(z)=\Sigma'(z)+i\Sigma''(z)$. The result above agrees with \cite{Aldape2022}.

   To observe quantum oscillations in $1/\omega_c$, we can apply Poisson resummation to \eqref{eq:sigmaDC}, yielding
  \begin{equation}\label{}
    \sigma^{xx}_\text{DC}=2e^2 \int \frac{\rd z}{2\pi}\left(-n_F'(z)\right) \sum_{k=-\infty}^{\infty} e^{2\pi i\frac{k}{\omega_c}(\mu+z-\Sigma'(z))} e^{-\frac{2\pi |k\Sigma''(z)|}{\omega_c}}(-1)^{k}\frac{-\Sigma''(z)}{\omega_c^2+4\left[\Sigma''(z)\right]^2}\left[z+\mu-\Sigma'(z)\right]\,.
  \end{equation} Since $z$ is integrated as in Sec.~\ref{sec:gpmag}, we expect the quantum oscillation to be similar to a Fermi liquid.

  The Hall conductivity is
  \begin{equation}\label{eq:sigmaxy}
    \sigma^{xy}_\text{DC}=2\Im \sigma_{\text{DC}}^{-+}=\frac{e^2 \omega_c^2}{2\pi}\int\frac{\rd z_1 \rd z_2}{(2\pi)^2}\mathcal{P}\frac{n_F(z_1)-n_F(z_2)}{(z_1-z_2)^2}\sum_n (n+1)A_{n+1}(z_1)A_n(z_2)\,.
  \end{equation}

 \subsubsection{Cyclotron Resonance }

 We can also compute Eq.\eqref{eq:sigmaxx} at finite frequency $\Omega$, yielding
 \begin{equation}\label{eq:sigmaxxOmega}
 \begin{split}
   \Re\sigma^{-+}(\Omega)&=\frac{e^2}{8\pi}\int \frac{\rd z}{2\pi}\frac{n_F(z)-n_F(z+\Omega)}{\Omega}\\
   &\Bigg[\frac{z+\mu-\Sigma(z)+\omega_c/2}{\left(\Omega-\omega_c-\Sigma(z+\Omega)+\Sigma(z)\right)}\left(\psi\left(\frac{3}{2}-\frac{z+\Omega+\mu-\Sigma(z+\Omega)}{\omega_c}\right)-\psi\left(\frac{1}{2}-\frac{z+\mu-\Sigma(z)}{\omega_c}\right)\right)\\
   &+\frac{z+\mu-\Sigma(z)+\omega_c/2}{\Omega-\omega_c-\bar{\Sigma}(z+\Omega)+\Sigma(z)}\left(\psi\left(\frac{1}{2}-\frac{z+\mu-\Sigma(z)}{\omega_c}\right)-\psi\left(\frac{3}{2}-\frac{z+\Omega+\mu-\bar{\Sigma}(z+\Omega)}{\omega_c}\right)\right)+c.c.\Bigg]\,.
 \end{split}
 \end{equation} Here $\sigma^{-+}(\Omega)$ only contains cyclotron resonances at positive frequencies. Here $\Sigma(z)$ is the retarded self-energy and $\bar{\Sigma}(z)$ is its complex conjugate.

\begin{figure}[htb ]
   \centering
   \includegraphics[width=0.75\textwidth]{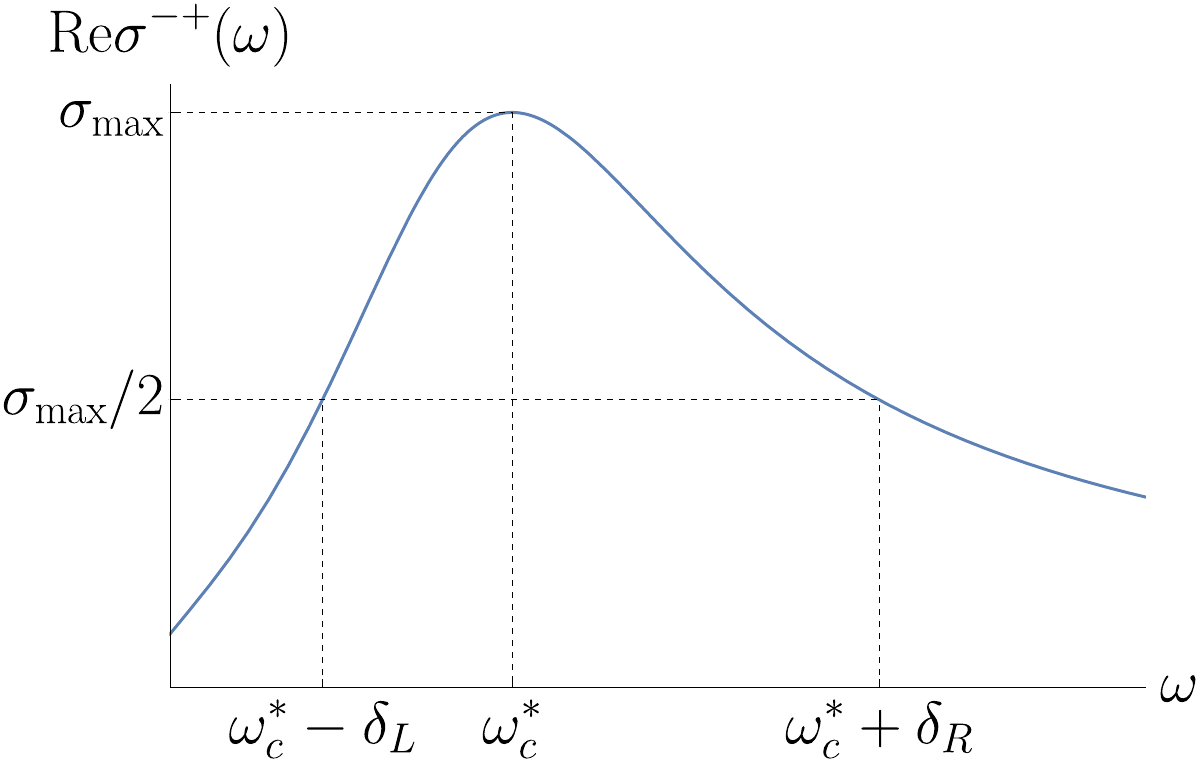}
   \caption{Example of the resonant optical conductivity $\Re\sigma^{-+}$ plotted as a function of frequency $\omega$. The renormalized cyclotron frequency $\omega_c^*$ is defined as the maximum of $\Re\sigma^{-+}$. We also defined $\delta_{L/R}$ to be the half-strength width on the left/right side of the peak.}\label{fig:exampleplt}
 \end{figure}

 To visualize the behavior of Eq.\eqref{eq:sigmaxxOmega}, we can set $T=0$ and substitute a MFL self-energy plus disorder scattering
 $$
 \Sigma(\omega)=A\left(\omega \ln\frac{|\omega|}{\omega_0}-i\frac{\pi}{2}|\omega|\right)-\frac{i \Gamma}{2}\,.
 $$ Here $A\propto g'^2$ quantifies the strength of interaction disorder and $\Gamma$ represents the scattering rate due to potential disorder. We also assume $\omega_c$ is not too large to allow for neglecting oscillatory terms in $\Sigma_R$ as discuused in Sec.~\ref{sec:gpnumres}.
 We next numerically evaluate the integral \eqref{eq:sigmaxxOmega}, and an example result is shown in Fig.~\ref{fig:exampleplt}. From Fig.~\ref{fig:exampleplt} we can define the renormalized cyclotron frequency $\omega_c^*$ as well as the half-strength widths $\delta_R$ and $\delta_L$.

\begin{figure}[htb ]
   \centering
   \includegraphics[width=0.75\textwidth]{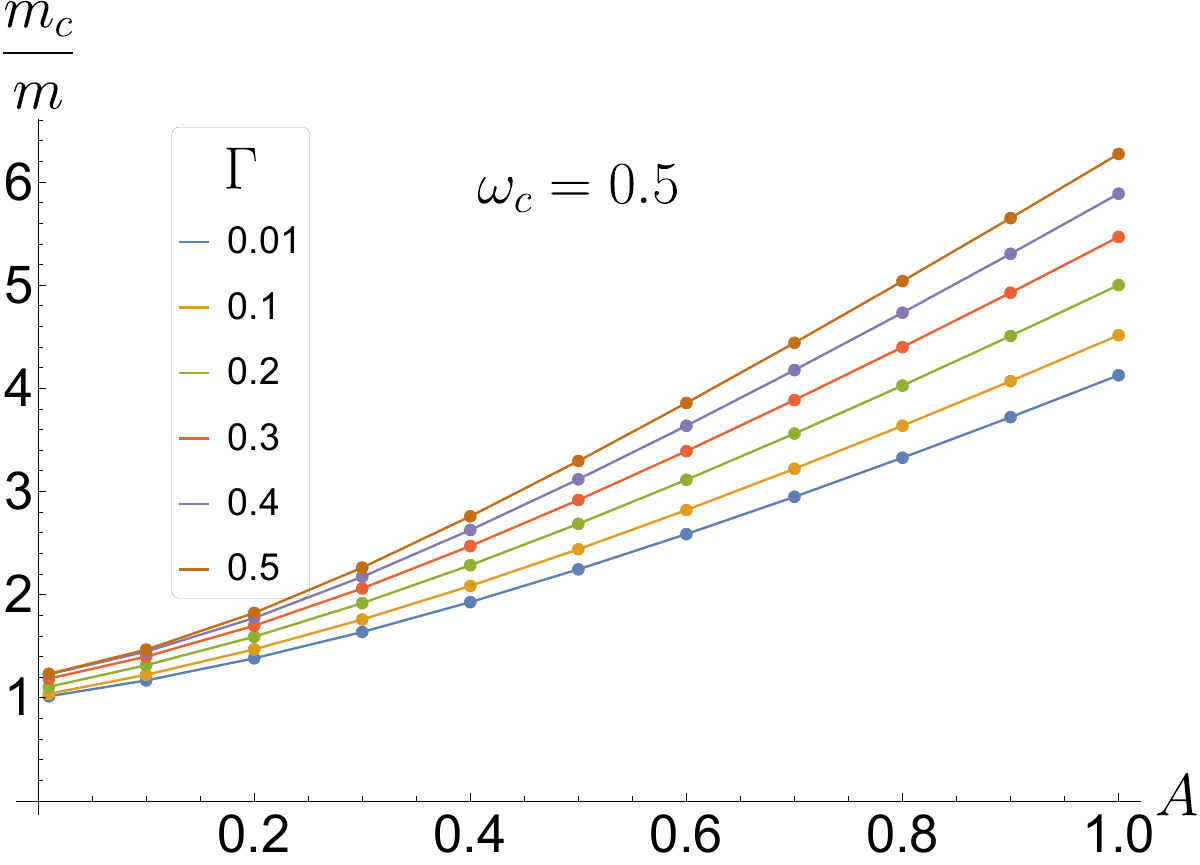}
   \caption{Cyclotron mass renormalization $\frac{m_c}{m}\equiv \frac{\omega_c}{\omega_c^*}$ at different values of potential disorder (represented by scattering rate $\Gamma$) and interaction disorder (represented by $A$). The bare cyclotron frequency is $\omega_c=0.5$. When $\omega_c>\Gamma$, the cyclotron mass is only weakly renormalized by $\Gamma$, but is more sensitive to $A$.}\label{fig:mcplt}
 \end{figure}

 We first present the result in terms of the cyclotron mass
 \begin{equation}\label{}
   \frac{m_c(\omega_c,A)}{m}\equiv \frac{\omega_c}{\omega_c^{*}}\,,
 \end{equation} which is plotted in Fig.~\ref{fig:mcplt} at fixed $\omega_c=0.5$ and various values of $A$ and $\Gamma$.
    We see that in the regime $\omega_c>\Gamma$, the disorder elastic scattering rate $\Gamma$ only weakly increases the cyclotron mass, but the disordered interaction on the other way is more effective.

\begin{figure}[htb ]
   \centering
   \includegraphics[width=0.75\textwidth]{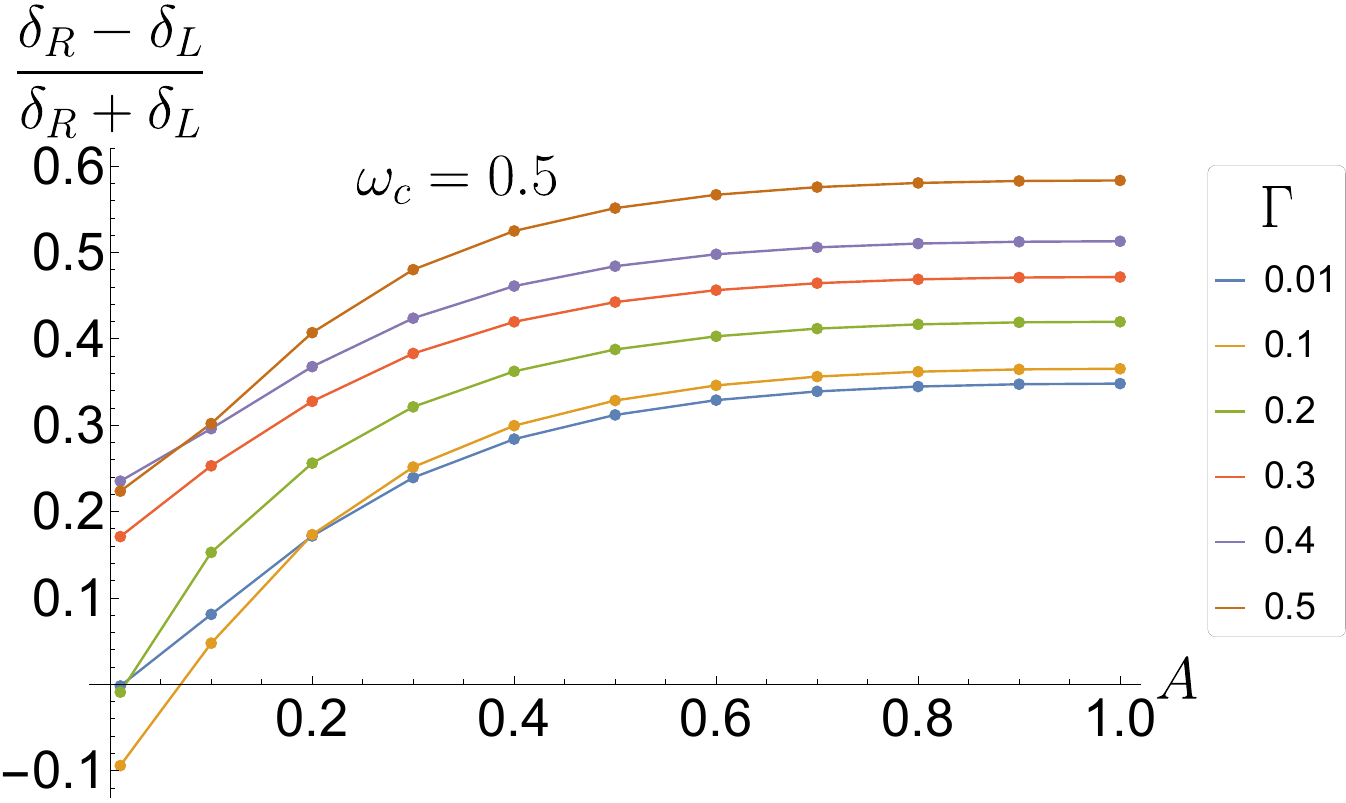}
   \caption{The asymmetry of the cyclotron resonant peak $\frac{\delta_R-\delta_L}{\delta_R+\delta_L}$ at different values of potential disorder (represented by scattering rate $\Gamma$) and interaction disorder (represented by $A$). The bare cyclotron frequency is $\omega_c=0.5$. The interaction disorder $A$ tends to make the resonant peak more asymmetric than the potential disorder $\Gamma$ does. }\label{fig:asymplt}
 \end{figure}

\begin{figure}[htb ]
   \centering
   \includegraphics[width=0.75\textwidth]{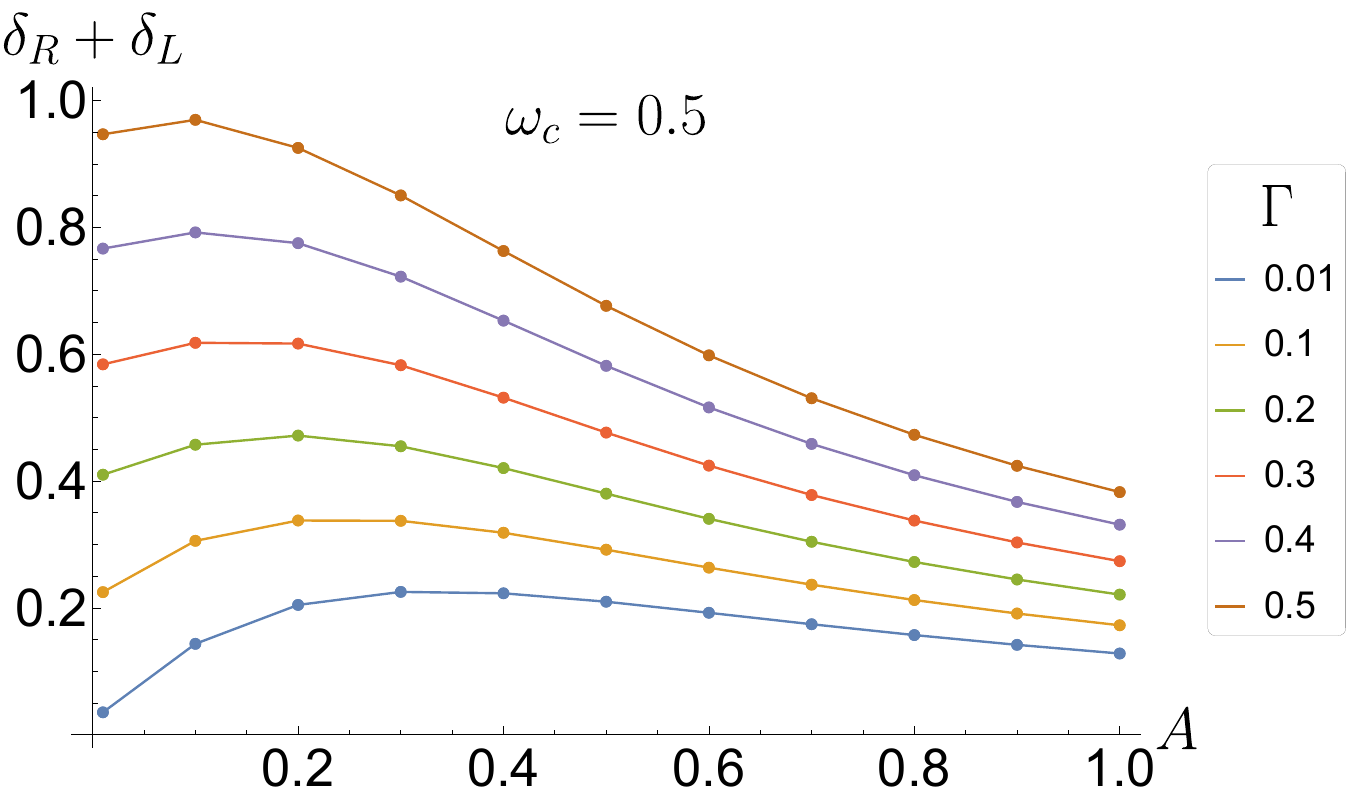}
   \caption{The total width of the cyclotron resonant peak $\delta_R+\delta_L$ at different values of potential disorder (represented by scattering rate $\Gamma$) and interaction disorder (represented by $A$). Potential disorder $\Gamma$ always make the resonant peak wider, but sufficiently large interaction disorder $A$ can in contrast sharpen the peak. Note that in the small $A$ limit the toal width is close to $2\Gamma$.}\label{fig:widthplt}
 \end{figure}

\begin{figure}
   \centering
   \includegraphics[width=0.75\textwidth]{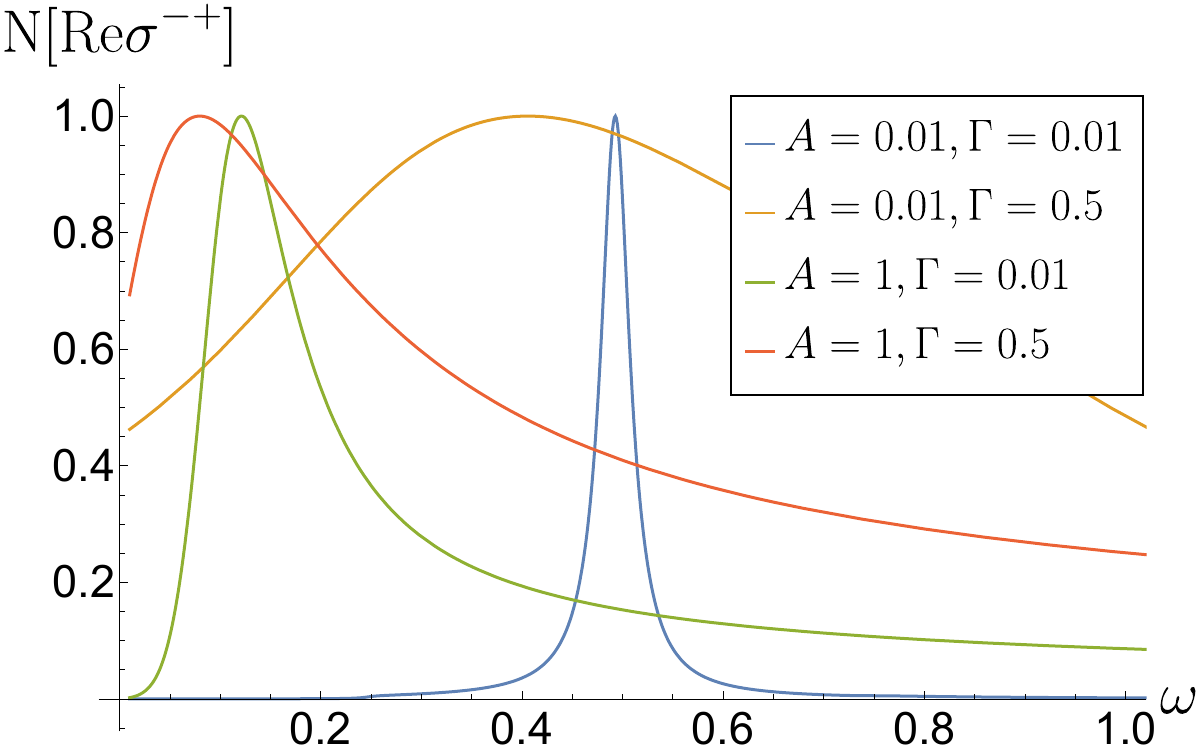}
   \caption{Comparison of normalized optical conductivity curves at different $(\Gamma,A)$. Here the bare cyclotron frequency is $\omega_c=0.5$ and the maxima have been normalized to unity.}\label{fig:cmpplt}
 \end{figure}

    In Figs.~\ref{fig:asymplt}-\ref{fig:cmpplt}, we show how the shape of the resonant peak evolve as we change $A$ and $\Gamma$ at fixed $\omega_c$. In Fig.~\ref{fig:asymplt}, we show that the interaction disorder $A$ increases the asymmetry of the resonant peak while the potential disorder $\Gamma$ results in a more symmetric peak. In Fig.~\ref{fig:widthplt}, we see that potential disorder $\Gamma$ always increase the peak width but sufficiently large interaction disorder $A$ can sharpen the peak. In Fig.~\ref{fig:cmpplt} we illustrate how the peak shape evolves as we change $(\Gamma,A)$.


\begin{figure}
        \centering
        \begin{subfigure}[b]{0.65\textwidth}
        \includegraphics[width=\textwidth]{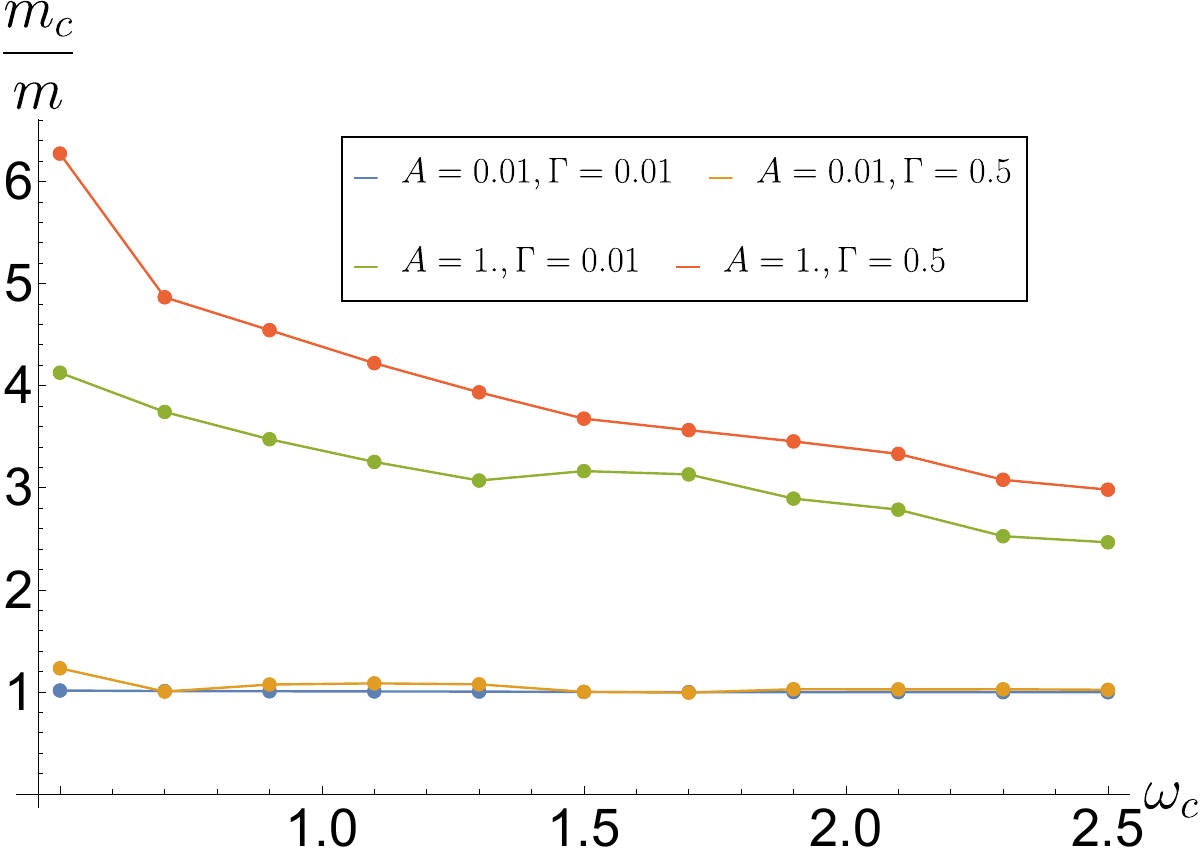}
        \caption{Cyclotron mass renormalization $\frac{m_c}{m}\equiv \frac{\omega_c}{\omega_c^*}$ as a function of the bare cyclotron frequency $\omega_c$ at different potential disorder scattering $\Gamma$ and interaction disorder $A$. We see that the interaction disorder introduces a substantial magnetic-field dependence in $m_c/m$ while the potential disorder does not. }
        \label{fig:mcplt2a}
        \end{subfigure}
        \begin{subfigure}[b]{0.65\textwidth}
        \includegraphics[width=\textwidth]{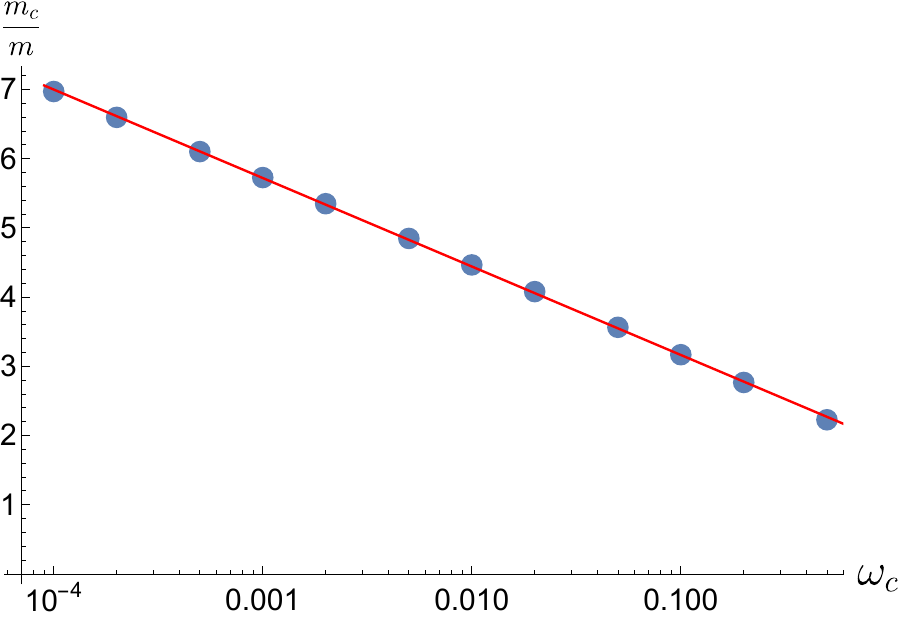}
        \caption{$m_c/m$ as a function of $\omega_c$ at fixed $A=0.5$ and $\Gamma=0.001$. The red line is the fitting to $\frac{m_c}{m}=a\ln\frac{|\omega_c|}{b}$.}\label{fig:ms_wc}
        \end{subfigure}
        \caption{}\label{fig:mcplt2}
    \end{figure}
\begin{figure}
        \centering
        \includegraphics[width=0.75\textwidth]{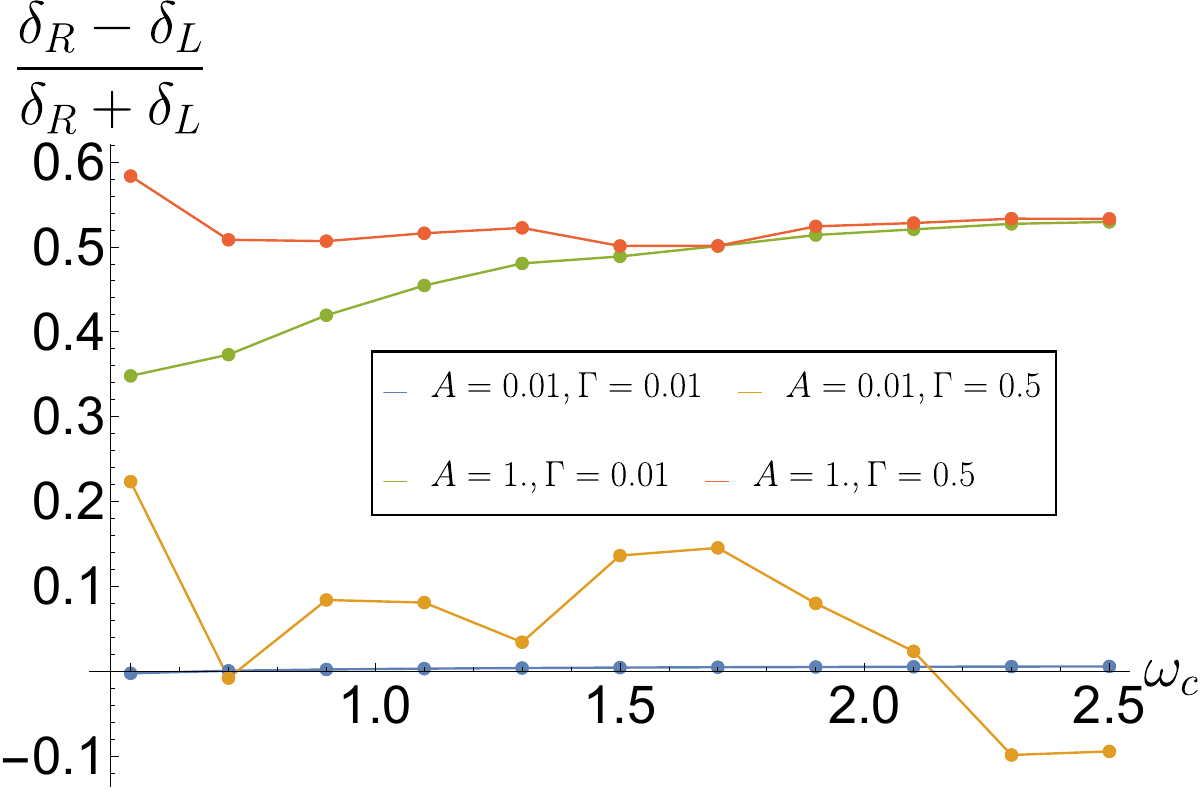}
        \caption{The asymmetry of the cyclotron resonant peak $\frac{\delta_R-\delta_L}{\delta_R+\delta_L}$ as a function of the bare cyclotron frequency $\omega_c$ at different potential disorder scattering $\Gamma$ and interaction disorder $A$. }
        \label{fig:asymplt2}
    \end{figure}
\begin{figure}
        \centering
        \includegraphics[width=0.75\textwidth]{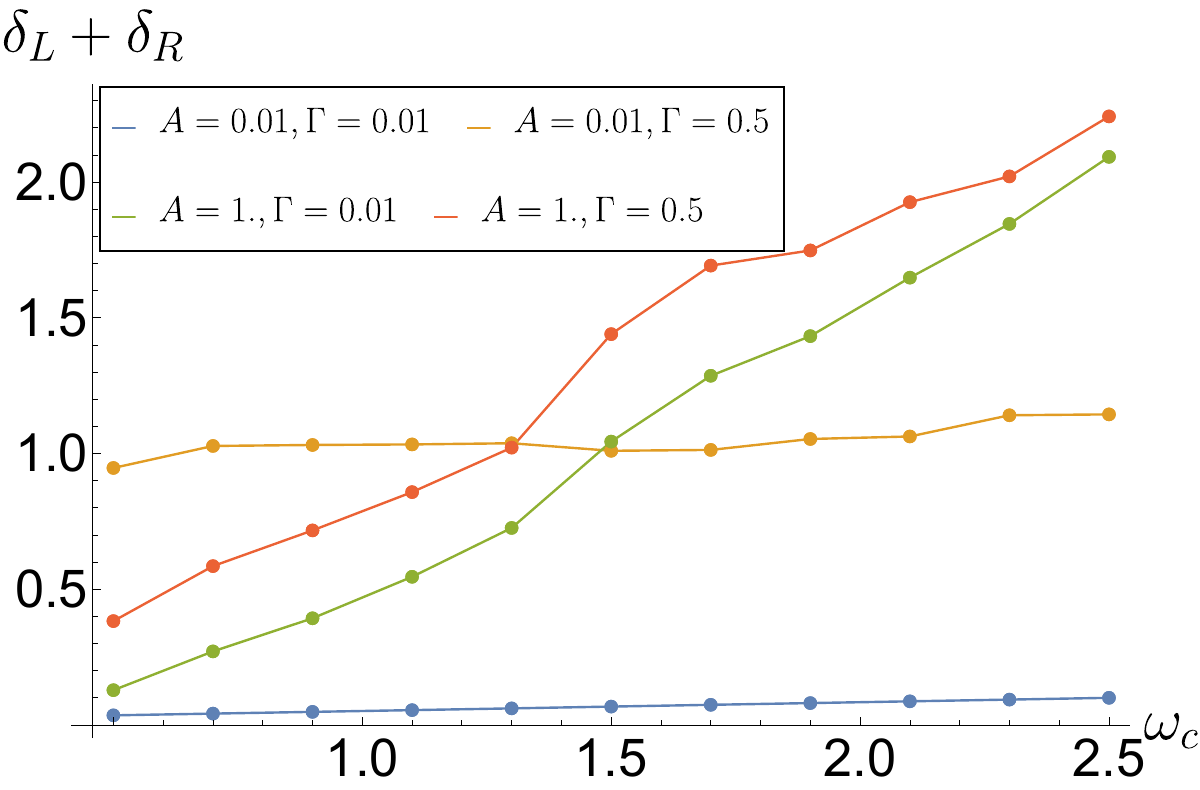}
        \caption{The total of the cyclotron resonant peak $\delta_R+\delta_L$ as a function of the bare cyclotron frequency $\omega_c$ at different potential disorder scattering $\Gamma$ and interaction disorder $A$. In the presence of disordered interaction, the width grows with the external magnetic field. }
        \label{fig:widthplt2}
    \end{figure}

    Next, we fix $\Gamma$ and $A$ and investigate the evolution of peak position and shape as we vary the external magnetic field (via varying the bare cyclotron frequency $\omega_c$). The results are shown in Figs.~\ref{fig:mcplt2}-\ref{fig:widthplt2}. In Fig.~\ref{fig:mcplt2}, we see that the interaction disorder induces a substantial magnetic field dependence in the cyclotron mass renormalization $m_c/m$. We also find that the asymmetry of the cyclotron peak only depends on $\omega_c$ weakly (Fig.~\ref{fig:asymplt2}), but the interaction disorder makes the peak width grow with $\omega_c$ (Fig.~\ref{fig:widthplt2}).

\begin{figure}[htb ]
   \centering
   \includegraphics[width=0.85\textwidth]{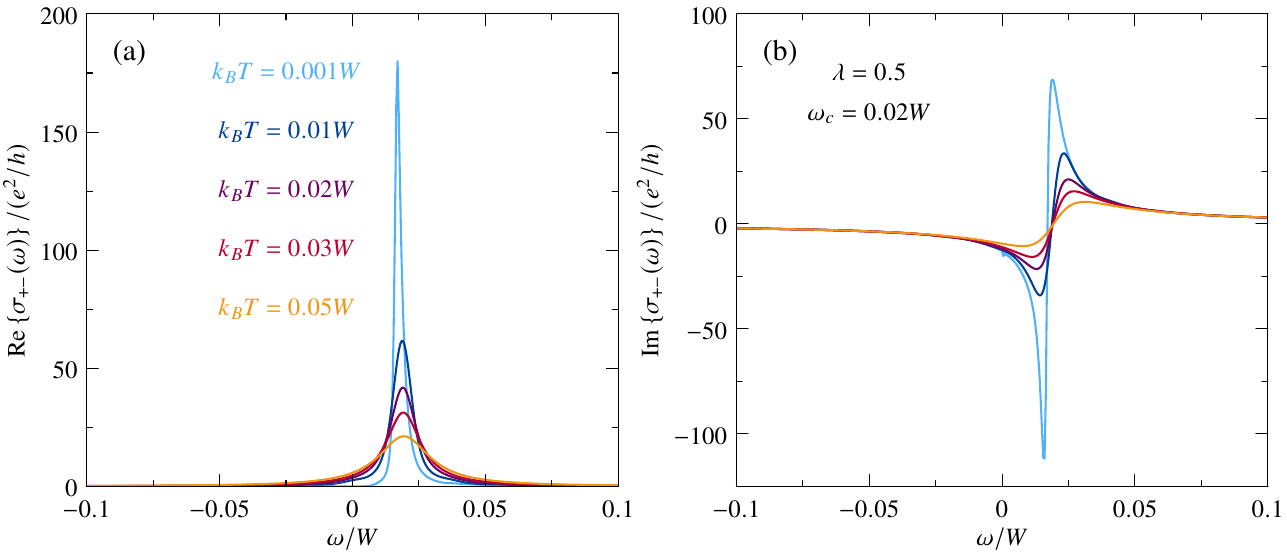}
   \caption{Real (a) and imaginary (b) part of the optical conductivity $\sigma_{+-}\left(\omega\right)$
as function of frequency for different temperatures at $\omega_{c}=0.02W$
and  for a dimensionless coupling constant $\lambda=0.5$.}
\label{fig:Fig_cond_lambda0p5}
 \end{figure}

\begin{figure}[htb ]
   \centering
   \includegraphics[width=0.85\textwidth]{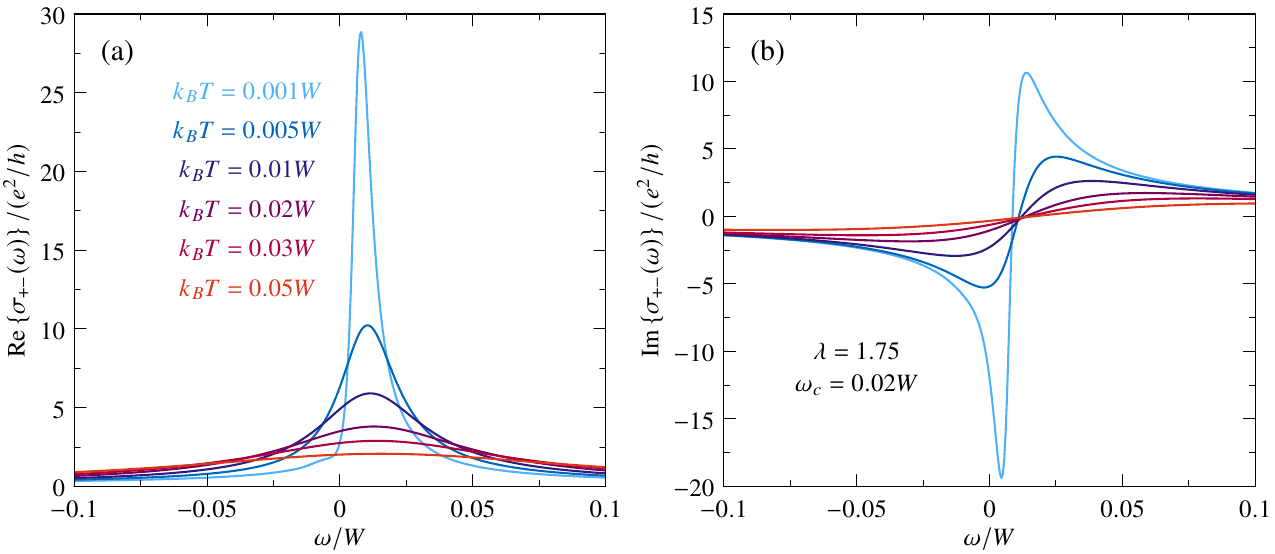}
   \caption{Real (a) and imaginary (b) part of the optical conductivity $\sigma_{+-}\left(\omega\right)$
as function of frequency for different temperture at $\omega_{c}=0.02W$
and dimensionless coupling constant $\lambda=1.75$. At this larger value of the coupling constant, the line-shape of the resonance shows a pronounced asymmetry, a behavior caused by the non-equally spaced Landau levels.}
\label{fig:Fig_cond_lambda1p75}
 \end{figure}

\begin{figure}[htb ]
   \centering
   \includegraphics[width=0.9\textwidth]{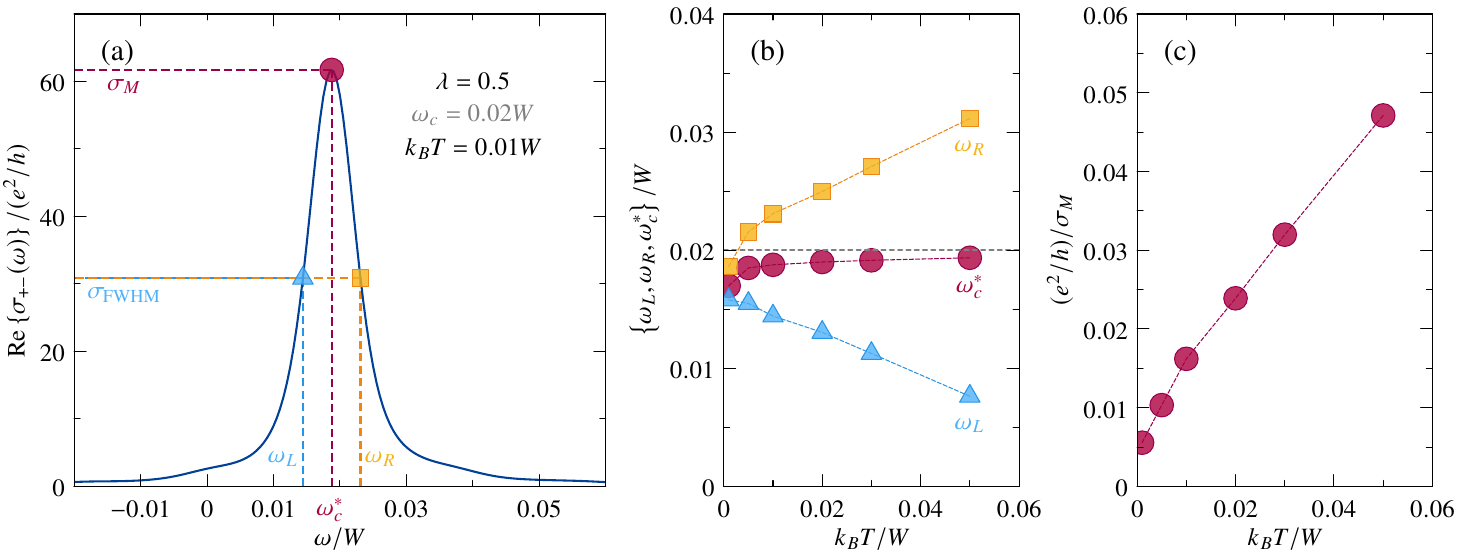}
   \caption{Analysis of the peak position $\omega_{c}^\ast$ and asymmetries $\omega_{R,L}=\omega_c^\ast\pm\delta_{R,L}$, introduced in panel (a), for a coupling constant $\lambda=0.5$. Panel (b) shows the temperature dependence
of the cyclotron mass and the peak width, while (c) shows the inverse
conductivity at the maximum.}
\label{fig:Fig_cond_extremes_lambda0p5}
 \end{figure}

\begin{figure}[htb ]
   \centering
   \includegraphics[width=0.9\textwidth]{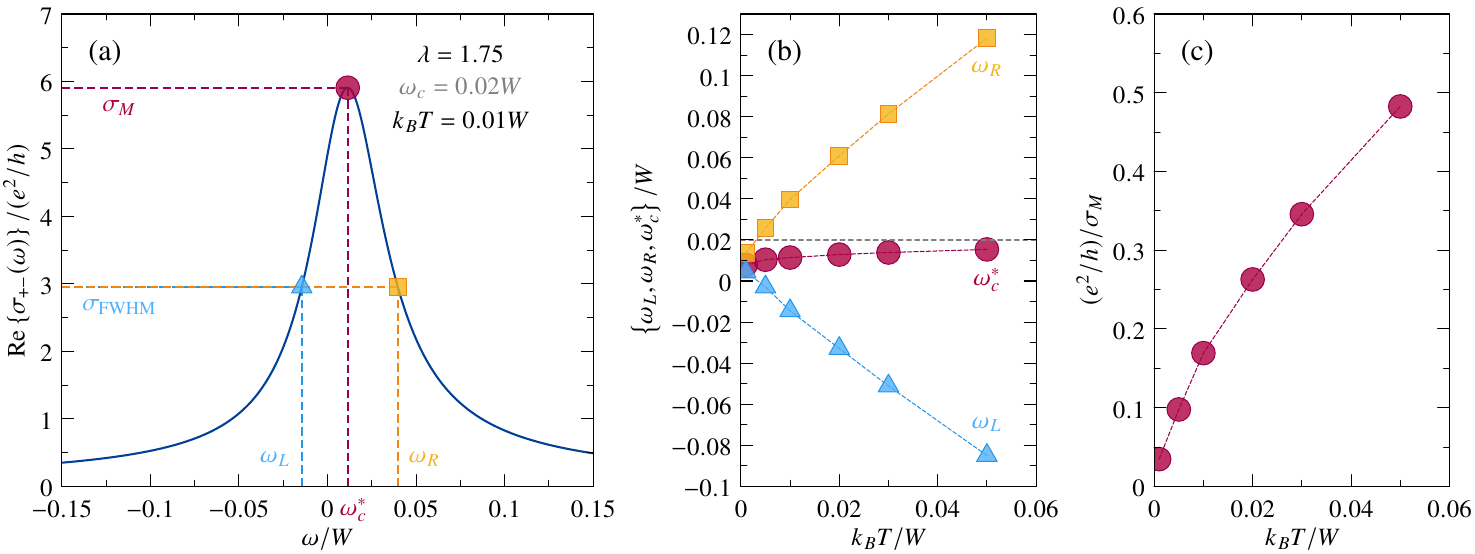}
   \caption{Analysis of the peak position $\omega_{c}^\ast$ and asymmetries $\omega_{R,L}=\omega_{c}^\ast\pm\delta_{R,L}$, introduced in panel (a), for a coupling constant $\lambda=1.75$.
 Panel (b) shows the temperature dependence
of the cyclotron mass and the peak width, while (c) shows the inverse
conductivity at the maximum.}
\label{fig:Fig_cond_extremes_lambda1p75}
 \end{figure}

\begin{figure}[htb ]
   \centering
   \includegraphics[width=0.9\textwidth]{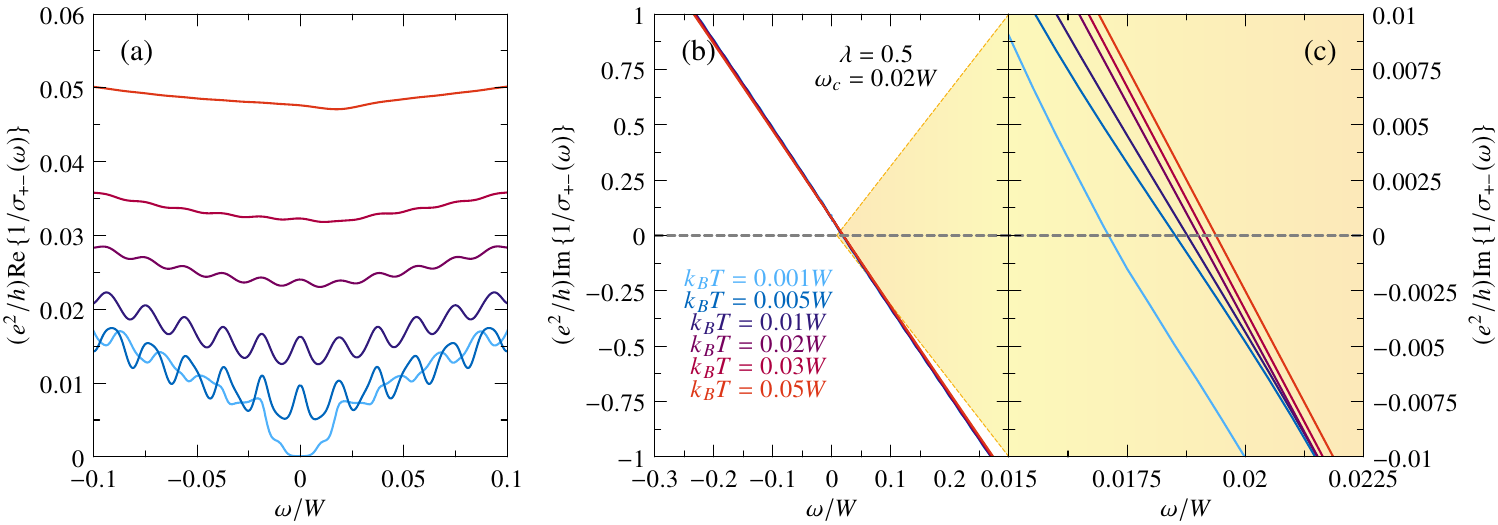}
   \caption{Real (a) and imaginary (b,c) part of the dynamical resistivity $1/\sigma_{+-}\left(\omega\right)$
as function of frequency for different temperatures at $\omega_{c}=0.02W$
and  for a dimensionless coupling constant $\lambda=0.5$. ${\rm Re}1/\sigma_{+-}(\omega)$ serves as a measure of the scattering rate while the imaginary part contains information about the effective mass and the cyclotron frequency. At this moderate coupling the zero crossing of ${\rm Im}1/\sigma_{+-}(\omega)$ and the peak position in ${\rm Re}\sigma_{+-}(\omega)$ coincide.}
\label{fig:Fig_Invcond_lambda0p5}
 \end{figure}

\begin{figure}[htb ]
   \centering
   \includegraphics[width=0.9\textwidth]{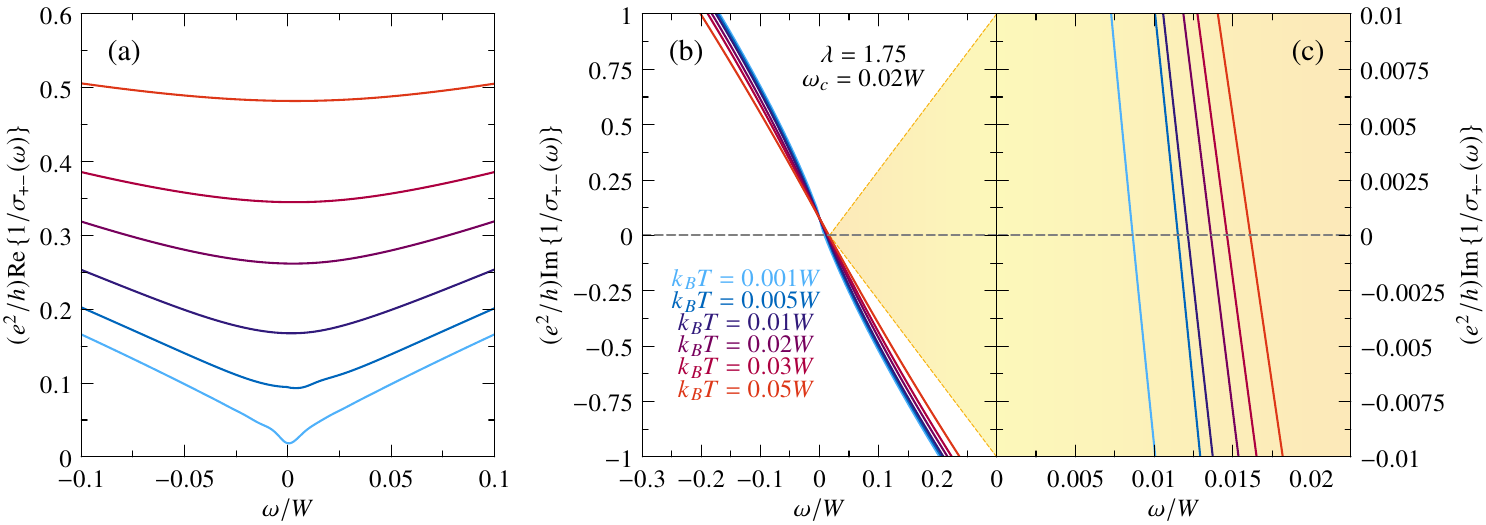}
   \caption{Real (a) and imaginary (b,c) part of the dynamical resistivity $1/\sigma_{+-}\left(\omega\right)$
as function of frequency for different temperatures at $\omega_{c}=0.02W$
and  for a dimensionless coupling constant $\lambda=1.75$. At this stronger coupling the zero crossing of ${\rm Im}1/\sigma_{+-}(\omega)$ and the peak position in ${\rm Re}\sigma_{+-}(\omega)$ are slightly different, indicating that the line shape affects the peak position. }
\label{fig:Fig_Invcond_lambda1p75}
 \end{figure}

 \joerg{Finally we determine the optical conductivity from the self-consistently determined self energy for different temperatures and coupling constants. The results for the real and imaginary parts of $\sigma_{+-}(\omega)$ are shown in Fig.~\ref{fig:Fig_cond_lambda0p5} for smaller values of the dimensionless coupling constant $\lambda=\frac{1}{2\pi}mg'^{2}=0.5$, and in Fig.~\ref{fig:Fig_cond_lambda1p75} for $\lambda=1.75$.  We observe a drift of the cyclotron resonance with temperature towards larger values as the temperature increases. In addition, the line shape of the resonance is asymmetric, a behavior particularly pronounced at strong coupling. The cyclotron resonance is a consequence of optical transitions between consecutive Landau levels. As was shown in Fig.~\ref{fig:Fig_DOS_extremes_lambda1p75}, at strong coupling, self energy renormalizations cause the distance between consecutive Landau levels to grow with energy scale, which is responsible for the asymmetric accumulation of weight around the cyclotron resonance. The line shape of the resonance can therefore be used to get a sense of the distribution of Landau level spacing. The change in the peak position and the changes of the widths of the resonance are show in Figs.~\ref{fig:Fig_cond_extremes_lambda0p5} and \ref{fig:Fig_cond_extremes_lambda1p75}, where we also show  the $T$-dependence of the inverse peak maximum. As the temperature increases the resonance position approaches the free-electron value, indicated by the dashed line in the panels (b) of the figures. The actual Landau levels, that, among others, enter the self energy are hardly visible in $\sigma_{+-}(\omega)$.
 To demonstrate that these effects are visible in the optical data themselves we plot in Figs.~\ref{fig:Fig_Invcond_lambda0p5} and \ref{fig:Fig_Invcond_lambda1p75} the real and imaginary parts of $1/\sigma_{+-}(\omega)$. At weak coupling one can clearly see the Landau levels in the frequency dependence of ${\rm Re}\left(1/\sigma_{+-}(\omega)\right)$ at lower temperatures, while they are strongly suppressed by the scattering at stronger coupling.
 For ${\rm Im}\left(1/\sigma_{+-}(\omega)\right)$ we find a zero crossing at the cyclotron frequency with significant changes in the slope at strong coupling. To illustrate this behavior in the simplifying Drude form}
 \begin{equation}
     \sigma_{+-}\left(\omega\right)=\frac{\omega_{p}^{2}}{4\pi}\frac{1}{\tau^{-1}+i\left(\omega_{c}-\omega\right)\left(1+\lambda\right)}
 \end{equation}
 \joerg{with scattering rate $\tau^{-1}$ and optical mass enhancement $\lambda$ one immediately finds
 ${\rm Re}\left(1/\sigma_{+-}\left(\omega\right)\right)\sim\tau^{-1}$ and ${\rm Im}\left(1/\sigma_{+-}\left(\omega\right)\right)\sim\left(1+\lambda\right)\left(\omega_{c}-\omega\right)$. The real part of the inverse susceptibility contains information about the scattering rate and the imaginary part about mass renormalization and the cyclotron resonance position. Comparing the cyclotron resonance energy obtained from the peak position of ${\rm Re}\left(\sigma_{+-}\left(\omega\right)\right)$ and the zero of ${\rm Im}\left(1/\sigma_{+-}\left(\omega\right)\right)$ they essentially agree at weak coupling but differ somewhat at stronger coupling, i.e. the large scattering rate affects the precise location of the cyclotron peak.  }

 \section{Translational Invariant $g$-model}\label{sec:gmodel}

    In supplement to the model with interaction disorder, in this part we study the translational invariant $g$-model and compare its properties to the $g'$-model.
\FloatBarrier
 \subsection{Action}

   The Hamiltonian for the translational invariant $g$-model is \cite{Esterlis2021,Guo2022,Patel2023}:
   \begin{equation}\label{eq:Sg0}
\begin{split}
   S =& \int \rd \tau \sum_{\vec{k}} \sum_{i=1}^{N}\psi_{i\vec{k}}^\dagger(\tau)\left[\partial_\tau+\varepsilon_{\vec{k}+\vec{A}}-\mu\right]\psi_{i\vec{k}}(\tau) \\
      +\frac{1}{2}&\int \rd \tau \sum_{i=1}^{N}\sum_{\vec{q}} \phi_{i,\vec{q}} \left[-\partial\tau^2+\omega_{\vec{q}}^2+m_b^2\right]\phi_{i,-\vec{q}}(\tau)\\
      +\frac{1}{N}&\int\rd^2 \vec{r}\rd \tau g_{ijl}\psi_i^\dagger(\vec{r},\tau)\psi_{j}(\vec{r},\tau)\phi_l(\vec{r},\tau)\,.
\end{split}
\end{equation} Here, the Yukawa coupling $g_{ijl}$ is translational invariant but are still randomly drawn from a Gaussian ensemble in flavor indices:
\begin{equation}\label{}
  \overline{g_{ijl}}=0\,,\qquad \overline{g^*_{ijl}g_{abc}}=\delta_{ia}\delta_{jb}\delta_{cl}\,.
\end{equation} The other quantities in the action \eqref{eq:Sg0} are defined in the same way as the $g'$-model.

    Performing average over $g_{ijl}$, we can derive the $G$-$\Sigma$ action in terms of bilinear fields:
    \begin{equation}\label{eq:SGSigma_conductivity}
\begin{split}
  \frac{1}{N}S
  =&-\ln\det((\partial_\tau+{\varepsilon}_{k+A}-\mu)\delta(x-x')+\Sigma)+\frac{1}{2}\ln\det\left(\left(-\partial_\tau^2+\omega_q^2+m_b^2\right)\delta(x-x')-\Pi\right)\\
  & -\int\rd^3 x\rd^3 x'\left(\Sigma(x',x)G(x,x')-\frac{1}{2}\Pi(x',x)D(x,x')\right) \\
       & +\int\rd^3 x\rd^3 x' \frac{g^2}{2}G(x,x')G(x',x)D(x,x')\,.
\end{split}
\end{equation}

    Taking functional derivatives of Eq.\eqref{eq:SGSigma_conductivity}, we obtain the following saddle point equations
\begin{eqnarray}
  G(x_1,x_2) &=& \left(-\partial_\tau+\mu-{\varepsilon}_{k+A}-\Sigma\right)^{-1}(x_1,x_2)\,,\label{eq:SD1} \\
  \Sigma(x_1,x_2) &=& g^2 G(x_1,x_2)D(x_1,x_2)\,,\label{eq:SD2} \\
  D(x_1,x_2) &=& \left(-\partial_\tau^2+\omega_q^2+m_b^2-\Pi\right)^{-1}(x_1,x_2)\,, \label{eq:SD3}\\
  \Pi(x_1,x_2) &=& -g^2 G(x_1,x_2)G(x_2,x_1)\,. \label{eq:SD4}
\end{eqnarray} Here the inverses in \eqref{eq:SD1} and \eqref{eq:SD3} are in functional sense. In the following, we will analyze Eqs.\eqref{eq:SD1}-\eqref{eq:SD4} in the case of uniform background magnetic field.

We continue to use the Landau level basis to evaluate the fermion Green's function.  The fermion Green's function, in a general operator form, can be written as
  \begin{equation}\label{}
    G(i\omega_n)=\frac{1}{i\omega_n+\mu-\Sigma(i\omega_n)-\hat{\varepsilon}_{k+A}}\,.
  \end{equation} Here we assume that the self-energy $\Sigma$ is only a function of frequency as in the zero field case \cite{Esterlis2021}, and therefore it commutes with the dispersion operator $\hat{\varepsilon}_{k+A}$. Unlike in the $g'$-model where $\Sigma(i\omega)$ is exactly proportional to the identity matrix in the Landau level basis, in the $g$-model we assume this only in the vicinity of the FS. Following this assumption, the Green's function can be expanded in the same way as Eq.\eqref{eq:GLandau} and Eq.\eqref{eq:Gxy}. In the following manipulations we assume that the typical value of the Landau level index $n$ to be vary large. As $n\omega_c$ should be comparable to $\mu=k_F^2/2m$, so $n\approx \nu_F=(k_F\ell_B)^2/2$.

\subsubsection{Boson self-energy}

  The boson self-energy from the $g$ coupling is
  \begin{equation}\label{}
    \Pi(x,y)=-g^2G(x,y)G(y,x)\,.
  \end{equation}

  The fourier transform is
  \begin{equation}\label{eq:Pi=GG}
    \Pi(i\Omega,q)=-\frac{g^2T}{2\pi\ell_B^2}\sum_{i\nu} \sum_{m,n} G_m(i\Omega+i\nu)G_n(i\nu) \underbrace{\int \frac{r\rd r\rd\theta}{\left(2\pi \ell_B^2\right)}\exp\left(-iqr\cos\theta-\frac{r^2}{2\ell_B^2}\right)L_n\left(\frac{r^2}{2\ell_B^2}\right)L_m\left(\frac{r^2}{2\ell_B^2}\right)}_{I_{m,n}(q\ell_B)}
  \end{equation}

  We evaluate the integral $I_{m,n}(z)$. First we perform the $\theta$-integral and rescale variables to obtain
  \begin{equation}\label{eq:Imn1}
    I_{m,n}(z)=\int_0^\infty t\rd t J_0(z t) e^{-\frac{t^2}{2}}L_n\left(\frac{t^2}{2}\right)L_m\left(\frac{t^2}{2}\right)\,,
  \end{equation} where $J_0$ is Bessel function. Next there are two ways to proceed which yield compatible results.

  \textbf{Method 1:}  This integral is tabulated in \cite{IntegralTableBook}, and the result is
  \begin{equation}\label{}
    I_{m,n}(z)=e^{-\frac{z^2}{2}}\left(-\right)^{m+n}L_n^{m-n}\left(\frac{z^2}{2}\right)L_m^{n-m}\left(\frac{z^2}{2}\right)\,,
  \end{equation} where $L_n^{\alpha}$ denote associated Laguerre polynomial.
  To proceed, we want to take the limit of large $n$ and $m$, i.e. $n=\nu_F+\Delta_n$ and $m=\nu_F+\Delta_m$ with $\Delta_{n(m)}$ being order one and $\nu_F\to \infty$. We use the asymptotic formula \cite{IntegralTableBook}
  \begin{equation}\label{eq:Lnasym}
    L_n^\alpha(x)=e^{\frac{x}{2}}\frac{n^{\alpha/2}}{x^{\frac{\alpha}{2}}}J_\alpha(2\sqrt{nx})\,,\quad n\to\infty\,,
  \end{equation} and obtain
  \begin{equation}\label{}
    I_{m,n}(z)=\left(\frac{n}{m}\right)^{\frac{m-n}{2}}J_{m-n}(\sqrt{2n}z)J_{m-n}\left(\sqrt{2m}z\right)\,.
  \end{equation} Next, we substitute the Bessel functions with asymptotics
  \begin{equation}\label{}
    J_\nu(z)\sim \sqrt{\frac{2}{\pi z}}\cos(z-\frac{1}{2}\pi\nu-\frac{1}{4}\pi)\,.
  \end{equation} Because $\sqrt{2m},\sqrt{2n}\approx k_F\ell_B$, the arguments of the Bessel functions are fast oscillating, and therefore we can replace $\cos^2$ by $1/2$ (the phase difference is $|\sqrt{2m}-\sqrt{2n}|q\ell_B\sim|\Delta_m-\Delta_n|q/k_F\ll 1$), and obtain
  \begin{equation}\label{eq:Imnused1}
    I_{m,n}(q\ell_B)=\frac{1}{\pi k_F q \ell_B^2}\,.
  \end{equation}

  \textbf{Method 2:}

  In Eq.\eqref{eq:Imn1}, we use Eq.\eqref{eq:Lnasym} first and then use the integral below tabulated in \cite{NIST:DLMF}
  \begin{equation}\label{eq:Imn2}
    I_{m,n}(z)=\int_0^\infty t\rd t J_0(z t)J_0(\sqrt{2n}t)J_0(\sqrt{2m} t)=\begin{cases}
                                                                              \frac{1}{2\pi A(\sqrt{2m},\sqrt{2n},z)}\,, & \mbox{if } |\sqrt{2m}-\sqrt{2n}|<z<\sqrt{2m}+\sqrt{2n} \\
                                                                              0, & \mbox{otherwise}.
                                                                            \end{cases}
  \end{equation} where $A(a,b,c)$ is the area of a triangle with sides $a,b,c$.

 In the large $k_F$ limit, $\sqrt{2m}=k_F\ell_B\left(1+\Delta_m/(k_F\ell_B)^2+\dots\right)$ and similarly for $\sqrt{2n}$.
  The lower bound of $z=q\ell_B$ in \eqref{eq:Imn2} is then $|\sqrt{2m}-\sqrt{2n}|=\frac{|\Delta_m-\Delta_n|}{k_F\ell_B}$. The lower bound is satisfied by the bosons in the critical regime and the weak-field limit. By weak field, we mean that the characteristic frequency is much larger than the cyclotron frequency $\omega\gg \omega_c$. A critical boson has $q_\omega\sim \gamma^{1/3}|\omega|^{1/3}\gg g^{2/3}v_F^{-1/3}|eB|^{1/3}$, and the second inequality follows from weak-field condition. To compare, the lower bound in \eqref{eq:Imn2} means $q_B\sim 1/(k_F\ell_B^2)\sim eB/k_F$, which is much smaller than $q_\omega$ in both power counting of $B$ and $1/k_F$. Furthermore, the scale of $q_B$ is also where the bosons start to feel the quantization of Landau levels.
  Therefore, in the kinematic regime of critical boson, we can ignore the magnetic field and use the leading order result of  \eqref{eq:Imn2}, which is
  \begin{equation}\label{eq:Imnused}
    I_{m,n}(q\ell_B)=\frac{1}{\pi q k_F \ell_B^2}\,.
  \end{equation}

  Using \eqref{eq:Imnused1} or \eqref{eq:Imnused},  we obtain
  \begin{equation}\label{}
    \Pi(i\Omega,q)=-\frac{\gamma Q(i\Omega)}{q}\,,
  \end{equation} where $\gamma=\frac{mg^2}{2\pi v_F}$ and
  \begin{equation}\label{eq:saddle_Pi}
    Q(i\Omega)= T\sum_{i\nu} \frac{\omega_c^2}{\pi^2}\sum_{m,n}G_m(i\nu+i\Omega)G_n(i\nu)\,.
  \end{equation}
  The sums over $m,n$ can be approximated to be from $-\infty$ to $\infty$, and we obtain
  \begin{equation}\label{eq:saddle_Pi_tan}
    Q(i\Omega)=\pi T\sum_{i\nu}\tan\left(\frac{\pi A(i\nu)}{\omega_c}\right)\tan\left(\frac{\pi A\left(i\nu+i\Omega\right)}{\omega_c}\right)\,,
  \end{equation}where $A(i\omega)=i\omega+\mu-\Sigma(i\omega)$.
  The $\omega_c\to 0$ limit of the above result can be recovered by noticing that the imaginary parts of the tangents become a sign and the real parts can be set to zero due to rapid oscillation, and therefore $\lim_{\omega_c\to 0} Q(i\Omega)-Q(0)=|\Omega|$.

  \subsubsection{Fermion self-energy}

  Now we calculate the Landau fermion self-energy, which in real space reads:
  \begin{equation}\label{}
    \Sigma(x,y)=g^2 G(x,y)D(x,y)\,.
  \end{equation}
  We now transform it to the Landau level basis
  \begin{equation}\label{eq:sigmanppnp}
  \begin{split}
    \Sigma_{n''k'',n'k'}(\tau)&=g^2 \sum_n \int\frac{\rd k}{2\pi} G_n(\tau)\int\frac{\rd^2\vec{q}}{(2\pi)^2}D(q,\tau)\\
    &\times \int\rd^2 \vec{y} e^{-i\vec{q}\vec{y}} \phi_{nk}^*(y)\phi_{n'k'}(y)
    \int \rd^2 x e^{i\vec{q}\vec{x}} \phi_{n''k''}^*(x)\phi_{nk}(x)\,.
  \end{split}
  \end{equation}
  The two integrals in the second line can be computed as
  \begin{equation}\label{eq:phiphi1}
    \int\rd^2 \vec{y} e^{-i\vec{q}\vec{y}} \phi_{nk}^*(y)\phi_{n'k'}(y)= 2\pi\delta(k-k'+q_1) e^{-ikq_2 \ell_B^2-\frac{i}{2}q_1q_2\ell_B^2-\frac{|q|^2\ell_B^2}{4}} \sqrt{\frac{n'!}{n!}}\left(\frac{\ell_B}{\sqrt{2}}(q_1-iq_2)\right)^{n-n'} L_{n'}^{n-n'}\left(\frac{|q^2|\ell_B^2}{2}\right)\,.
  \end{equation}
  \begin{equation}\label{eq:phiphi2}
     \int \rd^2 x e^{i\vec{q}\vec{x}} \phi_{n''k''}^*(x)\phi_{nk}(x)=2\pi \delta(k''-k-q_1)e^{ik'' q_2\ell_B^2-\frac{i}{2}q_1q_2\ell_B^2-\frac{|q|^2\ell_B^2}{4}}\sqrt{\frac{n!}{n''!}}\left(\frac{\ell_B}{\sqrt{2}}(-q_1+iq_2)\right)^{n''-n}
     L_n^{n''-n}\left(\frac{|q|^2\ell_B^2}{2}\right)\,.
  \end{equation} Using these and the assuming the boson propagator is circular symmetric in $q$, it's easy to verify that \eqref{eq:sigmanppnp} is nonzero only if $(n'k')=(n''k'')$. Therefore, the fermion self-energy is diagonal in the Landau level basis. We also see that the fermion self-energy is independent of $k$, which is expected because using $k$ is a gauge-dependent choice.

  The self-energy now becomes
  \begin{equation}\label{eq:Sigmank}
    \Sigma_{n'k'}(i\omega)=g^2 T\sum_{n,i\Omega} G_n(i\omega-i\Omega) \int\frac{\rd^2 q}{(2\pi)^2}D(q,i\Omega) e^{-\frac{|q|^2\ell_B^2}{2}}(-1)^{n-n'} L_{n'}^{n-n'}\left(\frac{q^2\ell_B^2}{2}\right)L_n^{n'-n}\left(\frac{q^2\ell_B^2}{2}\right)\,.
  \end{equation}

  To proceed, we use several properties of the associated Laguerre polynomials. The first is ($n,m$ are integers)
  \begin{equation}\label{eq:Ltransform}
    n! L_n^m(x)=(n+m)! L^{-m}_{n+m}(x)(-x)^{-m}\,.
  \end{equation}


  The second property is about the large $n$ asymptotic behaviors of $L_n^\alpha$ \cite{NIST:DLMF}:
  \begin{equation}\label{eq:Lasym}
    L_n^\alpha(x)=\frac{n^{\frac{\alpha}{2}-\frac{1}{4}}e^{\frac{x}{2}}}{\pi^{1/2} x^{\frac{\alpha}{2}+\frac{1}{4}}}\left[\cos\theta_n^\alpha(x)+\frac{b_1^\alpha(x)}{n^{1/2}}\sin\theta_n^\alpha(x)+\mathcal{O}\left(\frac{1}{n}\right)\right]\,,
  \end{equation}where
  \begin{eqnarray}
    \theta_n^\alpha(x) &=& 2\sqrt{nx}-\left(\frac{\alpha}{2}+\frac{1}{4}\right)\pi\,, \\
    b_1^\alpha(x) &=& =\frac{1}{48 x^{1/2}}(4x^2-12\alpha^2-24\alpha x-24x+3)\,.
  \end{eqnarray}This can also be alternatively written as
 \begin{equation}\label{eq:Lasym2}
   L_n^\alpha(x)=\frac{n^{\frac{\alpha}{2}-\frac{1}{4}}e^{\frac{x}{2}}}{\pi^{1/2} x^{\frac{\alpha}{2}+\frac{1}{4}}}
   \cos\left[\theta^\alpha_n(x)-\phi_n^\alpha(x)\right]\,,\quad \phi_n^\alpha(x)=\tan^{-1}\frac{b_1^\alpha(x)}{n^{1/2}}\,.
 \end{equation}

  Using the above formulas \eqref{eq:Ltransform}-\eqref{eq:Lasym2}, we obtain
  \begin{equation}\label{}
  \begin{split}
     e^{-\frac{|q|^2\ell_B^2}{2}}(-1)^{n-n'} L_{n'}^{n-n'}\left(\frac{q^2\ell_B^2}{2}\right)L_n^{n'-n}\left(\frac{q^2\ell_B^2}{2}\right) & =\frac{n'!(n')^{n-n'}}{n!\pi}\frac{2\cos^2\left[\theta_{n'}^{n-n'}\left(\frac{q^2\ell_B^2}{2}\right)\right]}{\sqrt{2n'q^2\ell_B^2}}=\frac{1}{\pi k_F q \ell_B^2}\,.
  \end{split}
  \end{equation}

   In the final result, we have used the fact that to leading order in $1/k_F$, we can replace $n'=n=k_F^2\ell_B^2/2$ and replace the oscillating part by its average. In this way we decoupled the Landau level summation and the boson momentum integral similar to the Prange-Kadanoff reduction \cite{Prange1964}.

   Now we can separately evaluate the $n$ summation and $q$ integral, obtaining a self-energy which is independent of the Landau level index:
   \begin{equation}\label{eq:Sigma=GD}
    \Sigma(i\omega)=g^2 T\int\frac{\rd^2\vec{q}}{(2\pi)^2}\sum_{i\Omega} \sum_n G_n(i\omega-i\Omega)\frac{D(\vec{q},i\Omega)}{\pi k_F|\vec{q}|\ell_B^2}\,.
  \end{equation}

  Eq.\eqref{eq:Sigma=GD} is also compatible with the zero-field limit. To see this, we can  simplify using $G_n(i\omega)^{-1}=A(i\omega)-\left(n+1/2\right)\omega_c$ and $D(q,i\Omega)^{-1}=q^2+\gamma Q(i\Omega)/q$, and obtain
  \begin{equation}\label{eq:saddle_Sigma}
    \Sigma(i\omega)=-\frac{g^2}{3\sqrt{3}v_F \gamma^{1/3}} T\sum_{i\Omega} \frac{\tan\left(\frac{\pi A(i\omega-i\Omega)}{\omega_c}\right)}{Q(\Omega)^{1/3}}\,.
  \end{equation} In the $\omega_c\to 0$ limit, $Q(i\Omega)\to|\Omega|$ and $\tan\left(\frac{\pi A(i\omega)}{\omega_c}\right)\to \sgn \omega$, and we obtain the result identical to \cite{Esterlis2021,Guo2022}.

    To summarize, we can simplify the saddle point equations of the clean model to the following
    \begin{eqnarray}
      G_n(i\omega) &=& \frac{1}{i\omega+\mu-\left(n+\frac{1}{2}\right)\omega_c-\Sigma(i\omega)} \\
      D(i\Omega,q) &=& \frac{1}{q^2+\gamma\frac{Q(i\Omega)}{q}} \\
      \Sigma(i\omega) &=& g^2 T\int\frac{\rd^2\vec{q}}{(2\pi)^2}\sum_{i\Omega} \sum_n G_n(i\omega-i\Omega)\frac{D(\vec{q},i\Omega)}{\pi k_F|\vec{q}|\ell_B^2} \\
      Q(i\Omega) &=& T\sum_{i\nu} \frac{\omega_c^2}{\pi^2}\sum_{m,n}G_m(i\nu+i\Omega)G_n(i\nu)
    \end{eqnarray}


   The main approximations involved are:
  (1) We used the $1/k_F$ expansion similar to the ones in the zero-field situations \cite{Guo2022} and should also be consistent with Prange-Kadanoff reduction \cite{Prange1964}.
  (2) In terms of response to magnetic field, there are mainly two categories: (a) Oscillatory response in fermion frequencies $A(i\omega)=i\omega+\mu-\Sigma(i\omega)$: this have been kept. (b) Oscillatory response in $q$: this have been ignored, because typical boson momentum $q$ is much larger than typical $A(i\omega)$ due to scaling so the oscillation is much faster.

\FloatBarrier
    \subsection{Numerical Solution for the Saddle Point Equations}

    \subsubsection{Method}

    In this part we solve the SD equations numerically to obtain the local density of states. Similar to the $g'$-model, we adapt the saddle point equations \eqref{eq:SD1}-\eqref{eq:SD4} to include appropriate UV cutoffs. The fermion Green's functions are handled in the same way as the $g'$-model, where we introduced the local Green's function $\bar{G}(i\omega)$ and $\bar{G}_R(\omega)$ as defined in Eq.\eqref{eq:SDpbarG} and Eq.\eqref{eq:SDpn1}.

     Next the boson self-energy is $\Pi(i\Omega,q)=-\gamma Q(i\Omega)/q$, where
    \begin{equation}\label{}
      Q(i\Omega)=\pi T \sum_{i\omega}\bar{G}_R(i\Omega+i\omega)\bar{G}(i\omega)\,.
    \end{equation} Analytically continue to real time, we obtain
    \begin{equation}\label{eq:SDn2}
      Q_R(t)=2\pi \Re\left[\bar{G}_R(t)\tilde{G}_R(t)^*\right]\,,
    \end{equation} where $\tilde{G}_R(\omega)=-2n_F(\omega)\Im \bar{G}_R(\omega)$. The fourier transform convention is $F(t)=\int\frac{\rd \omega}{2\pi} F(\omega)e^{-i\omega t}$\,.

     The fermion self-energy is
    \begin{equation}\label{eq:SDsigman}
      \Sigma(i\Omega)=-\frac{g^2}{v_F} T\sum_{i\Omega}\bar{G}(i\omega+i\Omega)\hat{D}(i\Omega)\,,
    \end{equation} where
    \begin{equation}\label{}
      \hat{D}(i\Omega)=\int_{q<\Lambda_q} \frac{\rd^2 q}{(2\pi)^2} \frac{1}{q}\frac{1}{\Omega^2+q^2-\frac{\gamma Q(i\Omega)}{q}}\,.
    \end{equation} The real-time version of $\hat{D}(i\Omega)$ is
    \begin{equation}\label{eq:SDn3}
      \hat{D}_R(\Omega)=\int_{q<\Lambda_q} \frac{\rd^2 q}{(2\pi)^2} \frac{1}{q}\frac{1}{\left(\eta-i\Omega\right)^2+q^2-\gamma Q_R(\Omega)/q}\,.
    \end{equation}

    Analytically continue Eq.\eqref{eq:SDsigman} to real time, we obtain
    \begin{equation}\label{eq:SDn4}
      \Sigma_R(t)=\frac{g^2}{v_F}\left[\bar{G}_R(t)\tilde{B}_{R}(t)^*-\tilde{G}_R(t)\hat{D}_R(t)\right]\,,
    \end{equation} where $\tilde{B}_R(\omega)=-2n_B(\omega)\Im \hat{D}_R(\omega)$.

    The numerical version of the Schwinger-Dyson equations are given by Eqs.\eqref{eq:SDpn1},\eqref{eq:SDn2},\eqref{eq:SDn3} and \eqref{eq:SDn4}.
    In the computation, we set the bandwidths $W/2=\Lambda_q=2$, and fixed the cyclotron frequency to be $\omega_c=0.1$, and also chose $\frac{m}{2\pi}=1$. The frequency axis is discretized to be a grid from $-\omega_\text{max}$ to $\omega_\text{max}=16$, with spacing $\Delta\omega=0.002$. The infinitesimal $\eta$ is chosen to be $\eta=10^{-4}$.

    \subsubsection{Numerical Results}

    Now we present the numerical results for the local density of states $\rho_\text{loc}(\omega)=\Im G_{\text{loc},R}(\omega)$, the imaginary parts of the retarded fermion self-energy $\Im \Sigma_R(\omega)$ and the imaginary parts of the boson damping function $\Im Q_R(\omega)$. In the zero-field limit we recover the $|\omega|^{2/3}$ scaling in $\Im\Sigma_R(\omega)$.

\begin{figure}[htb!]
      \centering
      \includegraphics[width=\textwidth]{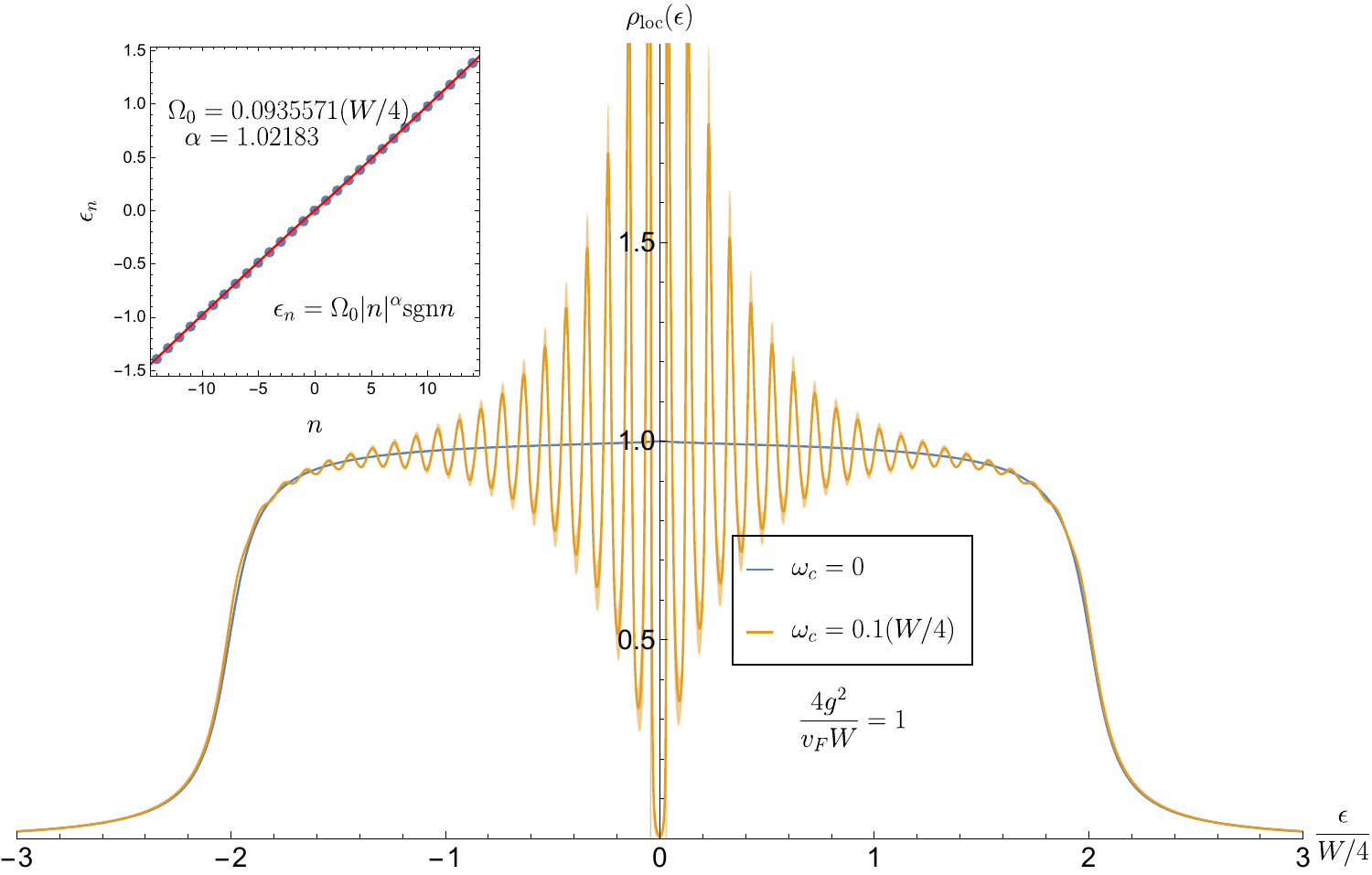}
      \caption{Local density of states $\rho_\text{loc}(\epsilon)=\Im G_{\text{loc},R}(\epsilon)$ at zero and nonzero magnetic fields (expressed in terms of cyclotron frequency $\omega_c$) and at coupling $4g^2/(v_FW)=1$.The inset plots the frequencies $\omega=\omega_n$ of local minima of $\rho_\text{loc}(\epsilon,\omega_c=0.1)$ as a function of index $n$. By fitting $\epsilon_n$ to a powerlaw in $n$, we see that $\epsilon_n$ is approximately linear in $n$ with spacing $\Omega_0$ close to the cyclotron frequency $\omega_c$. The shaded region represents numerical uncertainties.}\label{fig:coupling1_Gloc}
    \end{figure}

\begin{figure}[htb!]
      \centering
      \includegraphics[width=\textwidth]{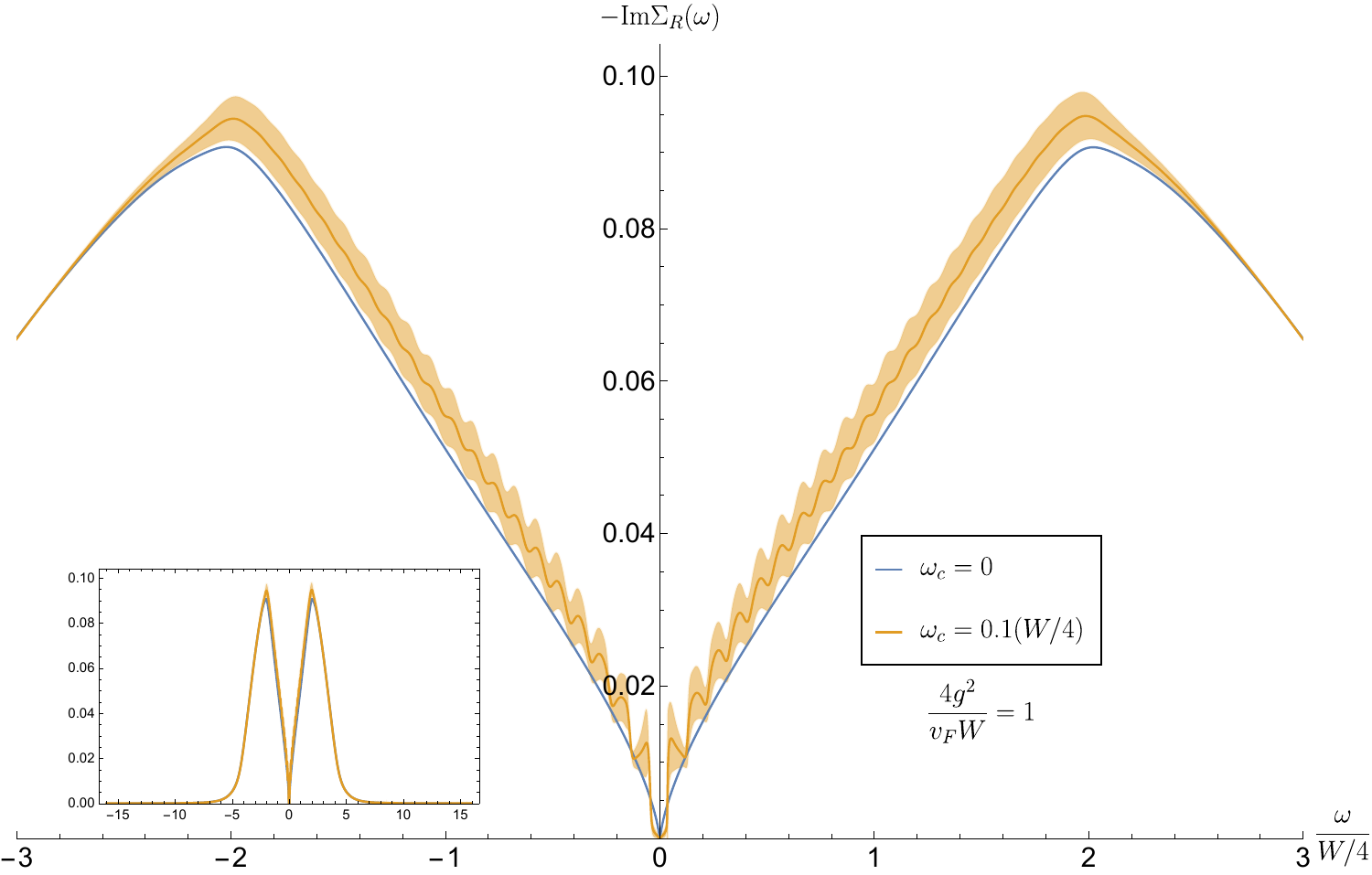}
      \caption{The imaginary part of retarded fermion self-energy $\Im\Sigma_R(\omega)$ at zero and nonzero magnetic fields (expressed in terms of cyclotron frequency $\omega_c$) and at coupling $4g^2/(v_FW)=1$. The inset plots the same function on a larger frequency range. The shaded region represents numerical uncertainties.}\label{fig:coupling1_Sigma}
    \end{figure}

\begin{figure}[htb!]
      \centering
      \includegraphics[width=\textwidth]{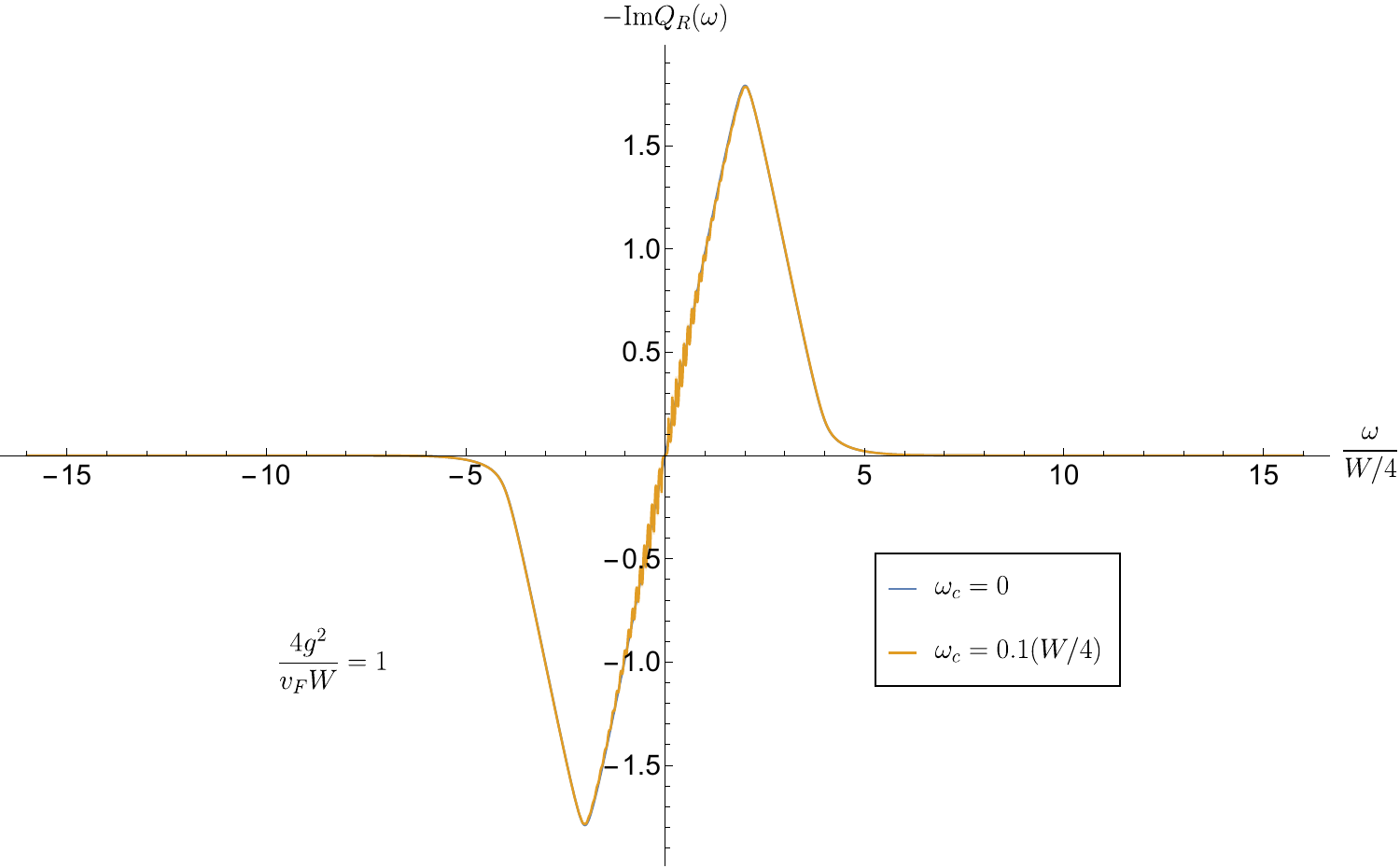}
      \caption{The imaginary part of retarded boson damping function $\Im Q_R(\omega)$ at zero and nonzero magnetic fields (expressed in terms of cyclotron frequency $\omega_c$).}\label{fig:coupling1_Q}
    \end{figure}

\begin{figure}[htb!]
      \centering
      \includegraphics[width=\textwidth]{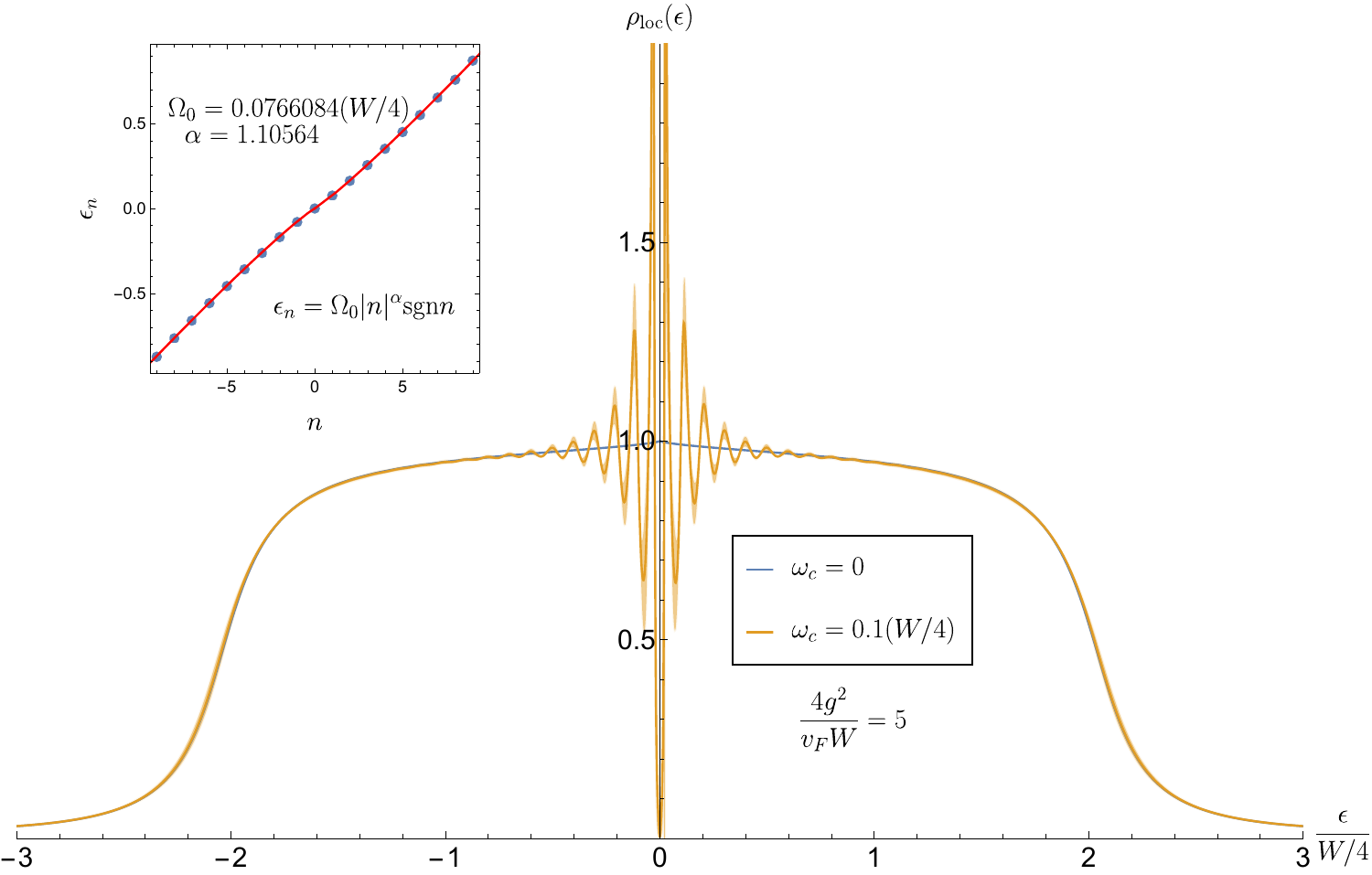}
      \caption{Same as Fig.~\ref{fig:coupling1_Gloc} but at coupling $4g^2/(v_FW)=5$.}\label{fig:coupling4_Gloc}
    \end{figure}

\begin{figure}[htb!]
      \centering
      \includegraphics[width=\textwidth]{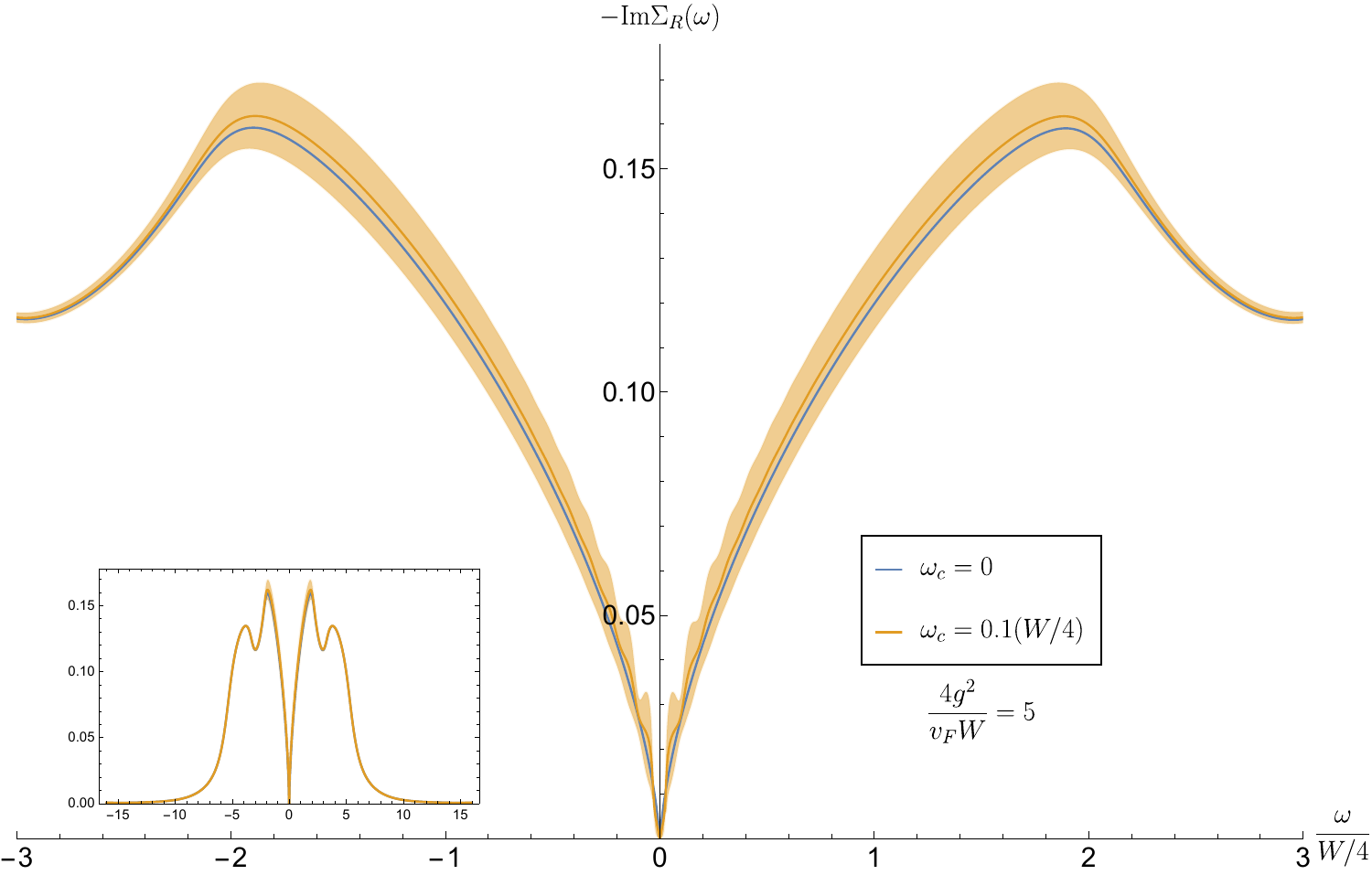}
      \caption{Same as Fig.~\ref{fig:coupling1_Sigma} but at coupling $4g^2/(v_FW)=5$.}\label{fig:coupling4_Sigma}
    \end{figure}

\begin{figure}[htb!]
      \centering
      \includegraphics[width=\textwidth]{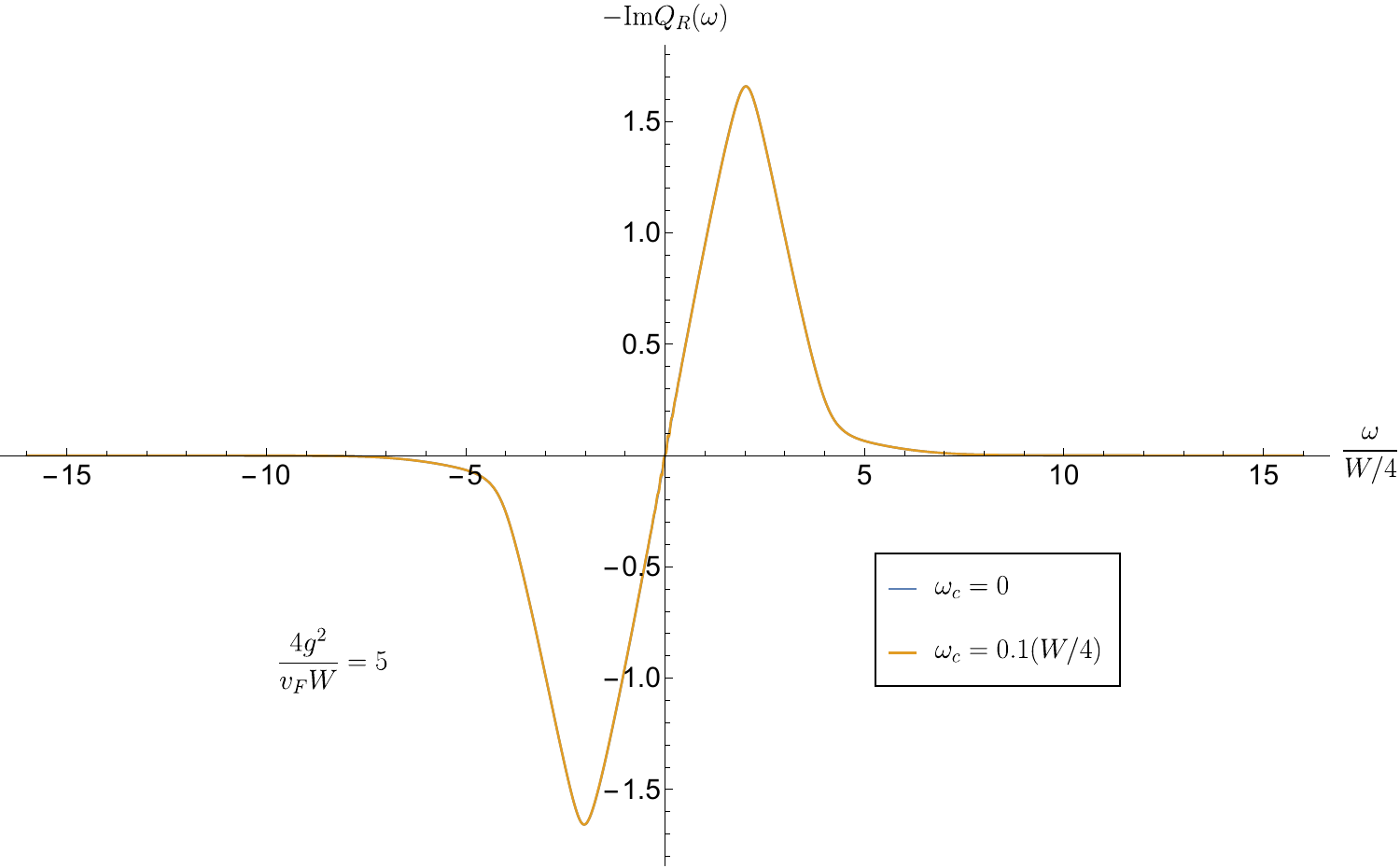}
      \caption{Same as Fig.~\ref{fig:coupling1_Q} but at coupling $4g^2/(v_FW)=5$.}\label{fig:coupling4_Q}
    \end{figure}

    We used two different couplings dimensionless couplings $4g^2/(v_F W)=1$ (Figs.~\ref{fig:coupling1_Gloc}-\ref{fig:coupling1_Q}) and $4g^2/(v_FW)=5$ (Fig.~\ref{fig:coupling4_Gloc}-\ref{fig:coupling4_Q}) to study how the oscillatory response vary with the strength of NFL interaction. The dichotomy between oscillatory behavior and NFL strength is similar to that of the $g'$-model.


    \FloatBarrier
    \subsection{Quantum oscillations in local density of states and magnetization}

      Since the local density of states and the magnetization is only sensitive to single-particle properties, the discussions in Secs.~\ref{sec:gpldos} and \ref{sec:gpmag} about the $g'$-model still apply here.
\FloatBarrier
      \subsection{Transport}
    In this part we compute the optical conductivity of the translational invariant $g$-model, and show that the cyclotron resonance is not renormalized by interaction, and re-establish Kohn's theorem. The framework of the computation is similar to Sec.~\ref{sec:gptransport}, but with different $W_\text{MT}$ and $W_\text{AL}$ as described below. Unlike the disordered interaction case where $W_\text{MT}$ and $W_\text{AL}$ cancel itself, here the two kernels should be treated carefully to account for drag effects.

  \subsubsection{MT diagram}

  The MT diagram kernel in real space reads
  \begin{equation}\label{eq:WMT}
    W_\text{MT}[F](x,y)=g^2D(x,y)F(x,y)\,.
  \end{equation} We transform this into the Landau level basis, writing
  \begin{equation}\label{eq:D_FT}
    D(x,y)=\int\frac{\rd \nu}{2\pi}\frac{\rd^2\vec{q}}{(2\pi)^2}e^{i\vec{q}\cdot(\vec{x}-\vec{y})-i\nu(\tau_x-\tau_y)}D(\vec{q},i\nu) \,,
  \end{equation}
  \begin{equation}\label{eq:F_FT}
    F(x,y)=\int\frac{\rd \Omega\rd \omega}{(2\pi)^2}e^{-i\omega(\tau_x-\tau_y)-i\Omega\frac{\tau_x+\tau_y}{2}}\sum_{n}\int\frac{\rd k}{2\pi}F_{n+1,n}(i\omega,i\Omega) \phi_{n+1,k}(\vec{x}) \phi_{n,k}^*(\vec{y})\,.
  \end{equation} In the following, we will show that the difference of Landau level indices $n'-n$ in $F_{n'n}$ is conserved (which is set to one explicitly because $F$ originates from the bare current vertex $\Gamma$), and $F_{n'n}$ can be consistently chosen as independent of $k$.

  We now substitute Eqs.\eqref{eq:D_FT}\eqref{eq:F_FT} into \eqref{eq:WMT}, and transform the spatial part into Landau level basis:
  \begin{equation}\label{eq:WMTF}
  \begin{split}
    W_\text{MT}[F]_{n'k',n''k''}(i\omega,i\Omega)&=g^2\sum_n\int\frac{\rd \omega'}{2\pi}\frac{\rd^2\vec{q}}{(2\pi)^2}\frac{\rd k}{2\pi}\rd^2\vec{x}\rd^2\vec{y}F_{n+1,n}(i\omega',i\Omega)D(\vec{q},i\omega-i\omega')e^{i\vec{q}\cdot(\vec{x}-\vec{y})}\\
    &\times \phi_{n'k'}^*(\vec{x}) \phi_{n+1,k}(\vec{x})\phi_{n''k''}(\vec{y})\phi^*_{nk}(\vec{y})\,.
  \end{split}
  \end{equation} The integrals over $\vec{x}$ and $\vec{y}$ can be evaluated analogously as in the previous section, yielding
  \begin{equation}\label{}
  \begin{split}
    \int\rd^2\vec{x}\phi_{n'k'}^*(\vec{x}) \phi_{n+1,k}(\vec{x}) e^{i\vec{q}\cdot\vec{x}}&=2\pi \delta(k+q_1-k')\exp\left(ik'q_2\ell_B^2-\frac{i}{2}q_1q_2\ell_B^2-\frac{|q|^2\ell_B^2}{4}\right)\sqrt{\frac{(n+1)!}{n'!}}\\
    &\times \left[\frac{\ell_B}{\sqrt{2}}(-q_1+iq_2)\right]^{n'-(n+1)} L_{n+1}^{n'-(n+1)}\left(\frac{|q|^2\ell_B^2}{2}\right)\,.
  \end{split}
  \end{equation}
  \begin{equation}\label{}
    \begin{split}
       \int\rd^2\vec{y}\phi_{n''k''}(\vec{y})\phi_{nk}^*(\vec{y})e^{-i\vec{q}\cdot\vec{y}} & =2\pi\delta(k+q_1-k'')\exp\left(-ikq_2\ell_B^2-\frac{i}{2}q_1q_2\ell_B^2-\frac{|q|^2\ell_B^2}{4}\right)\sqrt{\frac{n''!}{n!}} \\
         & \times \left[\frac{\ell_B}{\sqrt{2}}(q_1-iq_2)\right]^{n-n''} L_{n''}^{n-n''}\left(\frac{|q|^2\ell_B^2}{2}\right)\,.
    \end{split}
  \end{equation} These two integrals implies that in \eqref{eq:WMTF}  $k'=k''$ and the result is independent of $k'$. Because the remaining $\vec{q}$ integrals in \eqref{eq:WMTF} is rotationally symmetric, we must also have $n'=n''+1$, i.e. the difference is conserved.
  Combining everything together, we get
  \begin{equation}\label{eq:WMT2}
  \begin{split}
    W_\text{MT}[F]_{n''+1,n''}(i\omega,i\Omega)&=g^2\sum_n\int \frac{\rd \omega'}{2\pi}\frac{\rd^2\vec{q}}{(2\pi)^2}F_{n+1,n}(i\omega',i\Omega)D(\vec{q},i\omega-i\omega')\sqrt{\frac{n''+1}{n+1}}\\
    &\times\frac{n''!}{n!} L_{n''}^{n-n''}(z_q) L^{n-n''}_{n''+1}(z_q) z_q^{n-n''} e^{-z_q}\,,
  \end{split}
  \end{equation} where $z_q=\frac{|q|^2\ell_B^2}{2}$. Next, we use the asymptotic formula \eqref{eq:Lasym} for large $n,n''\approx \nu_F$, yielding
  \begin{equation}\label{}
  \begin{split}
    W_\text{MT}[F]_{n''+1,n''}(i\omega,i\Omega)&=g^2\sum_n\int \frac{\rd \omega'}{2\pi}\frac{\rd^2\vec{q}}{(2\pi)^2}F_{n+1,n}(i\omega',i\Omega)D(\vec{q},i\omega-i\omega')\\
    &\times\frac{2\cos\theta_{n''}^{n-n''}(z_q)\cos\theta_{n''+1}^{n-n''}(z_q)}{\pi k_F q \ell_B^2}\,.
  \end{split}
  \end{equation} As argued before, the oscillation of the $\cos$ factors in the critical region is fast, and therefore we can replace them by average
  \begin{equation}\label{}
    2\cos\theta_{n''}^{n-n''}(z_q)\cos\theta_{n''+1}^{n-n''}(z_q)\approx \cos\left(\theta_{n''}^{n-n''}(z_q)-\theta_{n''+1}^{n-n''}(z_q)\right)\approx\cos(\frac{q}{k_F})\,,
  \end{equation} which is exactly the scattering angle corresponding to MT diagram \cite{Guo2022}.

  We can now evaluate the $q$ integral using saddle point solutions, and expand the $\cos$ factor yielding
  \begin{equation}\label{eq:WMT3}
    W_\text{MT}[F]_{n''+1,n''}(i\omega,i\Omega)=g^2\sum_n\int \frac{\rd \omega'}{2\pi} F_{n+1,n}(i\omega',i\Omega)\frac{1}{\pi k_F\ell_B^2}\frac{1 }{3\sqrt{3}\gamma^{1/3}Q(\omega-\omega')^{1/3}}\left(1+\frac{\gamma^{2/3}Q(\omega-\omega')^{2/3}}{2k_F^2}\right)\,.
  \end{equation} Based on previous work \cite{Guo2022}, We expect that the first term in \eqref{eq:WMT3} should cancel with the self-energies, and the second term should cancel with AL diagrams.

  \subsubsection{AL diagram}

  In real space, the AL kernel reads
  \begin{equation}\label{}
    W_\text{AL}[F](x,y)=-g^4\int\rd^3 x' \rd^3 y' G(x,y)D(x,x')D(y',y)(F(x',y')G(y',x')+F(y',x') G(x',y'))\,.
  \end{equation} Switching to the Landau level basis, we obtain
  \begin{equation}\label{}
  \begin{split}
    W_\text{AL}[F]_{n',n''}(i\omega,i\Omega)=&\frac{1}{2\pi\ell_B^2}\sum_{n_1,n_2,n_3}\int\frac{\rd^2 \vec{q}}{2\pi}\frac{\rd \nu}{2\pi}\frac{\rd \omega'}{2\pi}D(\vec{q},i\nu+i\Omega/2)D(\vec{q},i\nu-i\Omega/2) e^{-2z_q} G_{n1}(i\omega-i\nu)\\
   & [G_{n2}(i\omega'-i\nu)-G_{n2}(i\omega'+i\nu)](-1)^{n'-n_1+n_3-n_2}\left[\frac{\ell_B}{\sqrt{2}}(q_1-iq_2)\right]^{n'-n''-1}\\
   &\sqrt{\frac{n''!}{n_1!}}L_{n''}^{n_1-n''}(z_q) \sqrt{\frac{n_1!}{n'!}}L_{n_1}^{n'-n_1}(z_q)\sqrt{\frac{(n_3+1)!}{n_2!}} L_{n_3+1}^{n_2-(n_3+1)}(z_q) \sqrt{\frac{n_2!}{n_3!}}L_{n_2}^{n_3-n_2}(z_q)\\
   & F_{n3+1,n_3}(i\omega',i\Omega)\,.
  \end{split}
  \end{equation} We see that the $q$ integral will enforce $n'=n''+1$.

  Next, we use \eqref{eq:Lasym2} to simplify the Laguerre polynomials, retaining terms that are slow varying in $q,n_1,n_2,n_3$, and to leading order in $|\vec{q}|/k_F$, we obtain
  \begin{equation}\label{}
  \begin{split}
    W_\text{AL}[F]_{n''+1,n''}(i\omega,i\Omega)&=-g^4\sum_{n_1,n_2,n_3}\int\frac{\rd^2 \vec{q}}{2\pi}\frac{\rd \nu}{2\pi}\frac{\rd \omega'}{2\pi}D(\vec{q},i\nu+i\Omega/2)D(\vec{q},i\nu-i\Omega/2)  G_{n1}(i\omega-i\nu)\\
    &[G_{n2}(i\omega'-i\nu)-G_{n2}(i\omega'+i\nu)]\frac{1}{4k_F^2\pi^3 |\vec{q}|^2 \ell_B^6}\left[1-\cos\frac{|\vec{q}|}{k_F}\right]F_{n_3+1,n_3}(i\omega',i\Omega)
  \end{split}
  \end{equation}

  \subsubsection{Resummation}

    Now we treat $W_\Sigma$. Similar to the zero-field case and follow the spirit of Prange-Kadanoff reduction \cite{Guo2022}, the Landau level indices of $F$ in $W_\text{MT}$ and $W_\text{AL}$ are summed over, and if we ignore the $n$-dependence in the bare verties (which is smooth),  the $W_\Sigma$ effectively becomes
    \begin{equation}\label{}
    \begin{split}
      L(i\omega,i\Omega)&=\sum_n W_{\Sigma,n+1,n}(i\omega,i\Omega)=\sum_n G_{n+1}(i\omega+i\Omega/2)G_n(i\omega-i\Omega/2)\\
      &=\underbrace{\frac{1}{i\Omega-\omega_c-\Sigma(i\omega+i\Omega/2)+\Sigma(i\omega-i\Omega/2)}}_{L_1(i\omega,i\Omega)}\underbrace{\frac{-\pi}{\omega_c}\left[-\tan\frac{\pi A(i\omega+i\Omega/2)}{\omega_c}+\tan\frac{\pi A(i\omega-i\Omega/2)}{\omega_c}\right]}_{L_2(i\omega,i\Omega)}
    \end{split}
    \end{equation}

    Now we have
    \begin{equation}\label{}
      \Pi^{-+}=\Gamma^{-,T}\frac{1}{W_\Sigma^{-1}-W_\text{MT}-W_\text{AL}}\Gamma^{+}=\Gamma^{-,T}L_2\frac{1}{L_1^{-1}-(W_\text{MT}+W_\text{AL})L_2}\Gamma^{+}\,.
    \end{equation} Since the bare vertices $\Gamma^{\pm}$ are frequency independent, we can now evaluate $L_{1}^{-1}$, $W_\text{MT}L_2$ and $W_\text{AL}L_2$ on a constant function:
    \begin{equation}\label{}
      L_{1}^{-1}[1](i\omega,i\Omega)=i\Omega-\omega_c-\Sigma(i\omega+i\Omega/2)+\Sigma(i\omega-i\Omega/2)\,.
    \end{equation}
    Notice that $L_2$ can be alternatively written as
    \begin{equation}\label{}
      L_2=\sum_n G_n(i\omega-i\Omega/2)-\sum_n G_n(i\omega+i\Omega/2)\,.
    \end{equation} We have
    \begin{equation}\label{eq:WMTL2}
      W_{\text{MT}}L_2[1](i\omega,i\Omega/2)=g^2\int \frac{\rd \omega'}{2\pi} \frac{\rd^2\vec{q}}{(2\pi)^2}D(\vec{q},i\omega-i\omega')\frac{\cos\frac{|\vec{q}|}{k_F}}{\pi k_F |\vec{q}|\ell_B^2}\sum_n\left[G_n(i\omega-i\Omega/2)-G_n(i\omega+i\Omega/2)\right]\,.
    \end{equation}
    \begin{equation}\label{eq:WALL2}
    \begin{split}
      W_{\text{AL}}L_2[1](i\omega,i\Omega/2)&=-g^4\sum_{n_1,n_2,n_3}\int\frac{\rd^2 \vec{q}}{2\pi}\frac{\rd \nu}{2\pi}\frac{\rd \omega'}{2\pi}D(\vec{q},i\nu+i\Omega/2)D(\vec{q},i\nu-i\Omega/2)  G_{n1}(i\omega-i\nu)\\
    &[G_{n2}(i\omega'-i\nu)-G_{n2}(i\omega'+i\nu)]\frac{1}{4k_F^2\pi^3 |\vec{q}|^2 \ell_B^6}\left[1-\cos\frac{|\vec{q}|}{k_F}\right]\\
    &\left[G_{n3}(i\omega-i\Omega/2)-G_{n3}(i\omega+i\Omega/2)\right]\,.
    \end{split}
    \end{equation} Now in \eqref{eq:WALL2}, we can compare the $n_2,n_3$ summation with saddle point equation \eqref{eq:Pi=GG} for $\Pi$, and using the identity
    \begin{equation}\label{}
      D(\vec{q},i\nu+i\Omega/2)D(\vec{q},i\nu-i\Omega/2)=\frac{1}{\Pi(\vec{q},i\nu+i\Omega/2)-\Pi(\vec{q},i\nu-i\Omega/2)}\left[D(\vec{q},i\nu+i\Omega/2)-D(\vec{q},i\nu-i\Omega/2)\right]\,,
    \end{equation}we obtain
    \begin{equation}\label{}
    \begin{split}
      W_\text{AL}L_2[1](i\omega,i\Omega)&=\int \frac{\rd \nu}{2\pi} \frac{\rd^2\vec{q}}{(2\pi)^2}\frac{g^2 }{\pi k_F|\vec{q}|\ell_B^2}\left[D(\vec{q},i\nu-i\Omega/2)-D(\vec{q},i\nu+i\Omega/2)\right]\\
      &\sum_{n_1}G_{n1}(i\omega-i\nu)
      \left[1-\cos\frac{|\vec{q}|}{k_F}\right]\,.
    \end{split}
    \end{equation} After some simple manipulations, we see that
    \begin{equation}\label{}
     W_\text{MT}L_2[1](i\omega,i\Omega)+W_\text{AL}L_2[1](i\omega,i\Omega)=\Sigma(i\omega-i\Omega/2)-\Sigma(i\omega+i\Omega/2)\,.
    \end{equation} Therefore
    \begin{equation}\label{}
      \frac{1}{L_1^{-1}-(W_\text{MT}+W_\text{AL})L_2}\Gamma^{+}=\frac{1}{i\Omega-\omega_c}\Gamma^{+}\,,
    \end{equation} meaning that the cyclotron frequency is not renormalized.

    We can continue to evaluate the current-current polarization, by substituting the expressions \eqref{eq:bareGammap},\eqref{eq:bareGammam} for the bare vertices
    \begin{equation}\label{}
    \begin{split}
      \Pi^{-+}(i\Omega)&=\frac{1}{i\Omega-\omega_c}\Gamma^{-,T}L_2 \Gamma^{+}\\
      &= \frac{1}{i\Omega-\omega_c}\frac{1}{2 m^2 \ell_B^2}\left[\int\frac{\rd k}{2\pi} 2\pi \delta(0)\right] \int\frac{\rd \omega}{2\pi} \sum_{n} (n+1)\left[G_n(i\omega-i\Omega/2)-G_n(i\omega+i\Omega/2)\right] \,.
    \end{split}
    \end{equation} The $k$-integral in the bracket can be evaluated using semiclassical state-counting arguments in textbooks, and the result is the Landau level degeneracy $\frac{S} {2\pi \ell_B^2}$ where $S$ is the system area. In evaluating the $n$-summation, we can set the factor $n+1=\nu_F=k_F^2\ell_B^2/2$ as the most contribution is from the Fermi surface. We then obtain the optical conductivity before analytic continuation to be
    \begin{equation}\label{}
      \sigma^{-+}(i\Omega)= \frac{1}{4}\calN v_F^2 \frac{1}{\Omega+i\omega_c} \calD(i\Omega,\omega_c)\,.
    \end{equation} Here $\calN=m/(2\pi)$ is the density of states on the FS at zero-field, and the function $\calD$ is
    \begin{equation}\label{eq:calD1}
      \calD(i\Omega,\omega_c)=\frac{i}{2\Omega}\int\rd \omega \frac{\omega_c}{\pi}\sum_{n}\left[G_n(i\omega+i\Omega/2)-G_n(i\omega-i\Omega/2)\right]\,,
    \end{equation}
    which describes the modulation of density of states due to magnetic field and finite frequency.  We now evaluate the function $\calD$ using Poisson resummation formula. Substituting $G_n(i\omega)^{-1}=A(i\omega)-\left(n+1/2\right)\omega_c$, and apply Poisson resummation to $n$, we obtain
    \begin{equation}\label{eq:calD2}
      \calD(i\Omega,\omega_c)=1+\frac{1}{\Omega}\sum_{k=1}^{\infty} \int \rd\omega(-1)^k \left[\calF_k(i\omega+i\Omega/2)-\calF_k(i\omega-i\Omega/2)\right].
    \end{equation} The function $\calF_k$ is
    \begin{equation}\label{eq:Fk}
      \calF_k(i\omega)=\theta(\omega)\exp\left(\frac{2\pi i k}{\omega_c}A(i\omega)\right)-\left(1-\theta(-\omega)\right)\exp\left(-\frac{2\pi i k}{\omega_c}A(i\omega)\right)\,.
    \end{equation} Because of causality $\sgn \Im A(i\omega)\propto \sgn \omega$, each term in \eqref{eq:Fk} is exponentially decaying as $\omega\to\pm \infty$, and therefore the integral $\int\rd \omega \calF_k(i\omega)$ converges. This implies that each term in the sum of Eq.\eqref{eq:calD2} vanishes, and therefore
    \begin{equation}\label{}
      \calD(i\Omega,\omega_c)=1\,.
    \end{equation}

    Therefore, the optical conductivity is exactly of Drude form
    \begin{equation}\label{}
      \sigma^{-+}(i\Omega\to \omega)=\frac{1}{4}\calN v_F^2 \frac{1}{-i\omega+i\omega_c}\,,
    \end{equation} and
    \begin{equation}\label{}
      \sigma^{xx}(\omega)=\sigma^{-+}(\omega)+\sigma^{+-}(\omega)=\frac{\calN v_F^2}{2}\frac{i \omega}{\omega^2-\omega_c^2}\,.
    \end{equation}

    Therefore, in the $g$-model the cyclotron frequency does not renormalize and there is no Shubnikov-de Haas oscillation.
\FloatBarrier
\subsection{Comment on the situation with translational invariant Yukawa coupling, potential disorder and interaction disorder}

  We briefly discuss the situation where translational invariant Yukawa coupling $g$, potential disorder $v$ and interaction disorder $g'$ are all present.  Because Prange-Kadanoff reduction breaks down in the presence of $v$ , the model can only be studied perturbatively in $g$ and $g'$ \cite{Guo2022}. Since both $v$ and $g'$ do not conserve momentum, they do not contribute to the vertex correction but only the fermion self-energy, and their effects can still be described by Sec.~\ref{sec:gpmodel}. As for the $g$-coupling, since translational invariance is broken we no longer expect exact cancellation between the Maki-Thompson and Aslamazov-Larkin diagrams so $g$ can potentially renormalize the cyclotron frequency.  In Ref.~\cite{Guo2022}, it is shown that due to the small-angle nature of the $g$-scattering, the self-energy due to $g$ is almost cancelled by vertex correction. The result of this cancellation is a renormalization of the bare $i\Omega$ term and a $\Omega^2$ transport scattering rate, both of which have a prefactor that vanishes in the $k_F\to\infty$ limit. We do not the expect this result to be altered by the weak magnetic field, therefore the effect of $g$ should be sub-extensive and negligible.

\section{Conclusion}

The properties of most metals in moderate magnetic field are well described by a free electron picture. This is justified by the principles of Fermi liquid theory, and Kohn's theorem \cite{Kohn1961,Kanki1997} which constrains the conductivity to be exactly that of free electrons in the presence of a Galilean translational symmetry. Of course, electrons in metals are not in a Galilean invariant environment, but nevertheless there is an emergent Galilean symmetry: as long as the physical properties are determined by long-wavelength interactions near the Fermi surface, it makes little difference if the regions far from the Fermi surface are not Galilean invariant. Only large momentum interactions, involving umklapp, can lead to renormalization of the cyclotron frequency from its free electron value \cite{Kanki1997}.

Turning our attention to non-Fermi liquids in clean metals, we can again expect Kohn's theorem to place strong constraints on the conductivity because the singular interactions are at long wavelengths. At zero magnetic field, this was highlighted in recent work \cite{Guo2022,ShiLoop}, showing that a proposed \cite{Kim94} optical conductivity $\sim 1/\omega^{2/3}$ was absent.
Section~\ref{sec:gmodel} on the $g$-model extends these analyses to non-zero magnetic field, and shows that Kohn's theorem continues to apply.

In the light of these theoretical studies, the recent observations by Legros {\it et al.} \cite{Legros2022} do appear quite surprising. In the present paper we show that the route to linear-in-$T$ resistivity of strange metals in Ref.~\cite{Patel2023} also provides a route to resolving the puzzles posed by cyclotron resonance. Specifically, a spatially random Yukawa coupling between the electrons and the critical boson leads to a full breakdown of Kohn's theorem. Our computations are described in Section~\ref{sec:gpmodel}, and the main results are in Figs~\ref{fig:exampleplt}-\ref{fig:cmpplt}.
These results show renormalization of the cyclotron mass and the lineshape as a function on the random interactions, and also random potential scattering.

We comment on the relation between the cyclotron mass and the thermodynamic mass (measured through heat capacity). In the $\Gamma=0$ limit, it can be shown numerically that the cyclotron mass is logarithmic in $\omega_c$, which follows from the MFL self-energy of the fermion. At the same time, the thermodynamic mass in the heat capacity also shows a $\ln T$ enhancement in the $T\to 0$ limit which is due to the $|\Omega|$ damping in the boson self-energy \cite{Esterlis2021}. In the pure $g'$ model, the coefficient of these two logarithmics are related because they are both functions of $g'$ and the fermionic density of states \cite{Esterlis2021}. However, for a general theory where the translational invariant coupling $g$ and potential disorder $v$ are present, these two logarithms are expected to be different. This is because the combination of $g$ and $v$ can also produce a $|\Omega|$ damping for the boson and contribute to the thermodynamic mass \cite{Guo2022}, but their effects cancel in the cyclotron resoannce.

We expect similar conclusions to those in this work to hold in the case of non-Fermi liquids arising near non-symmetry breaking Fermi volume changing transitions \cite{Senthil2004, Zhang2020, Aldape2022,ORE}. As long as the bosons mediating the hybridization between the conduction electrons and emergent fermions are uncondensed, the charge transport properties are dominated by the conduction electrons, with an inelastic scattering rate set by the spatially random ({\it i.e.} $g'$) part of the fermion-boson Yukawa coupling \cite{Aldape2022}. The contribution of the spatially uniform ({\it i.e.} $g$) part of the coupling to transport is weak as it only scatters electrons off-shell due to the Fermi surfaces of the electrons and the emergent fermions being generically unmatched. Therefore, most of the enhancement of the cyclotron mass in these situations should also arise from disorder effects.

Also of interest is non-Fermi liquid behavior that is restricted to small portions of the conduction electron Fermi surface (known as ``hot spots"), which arises from near-critical bosonic fluctuations centered at a finite wavevector, such as antiferromagnetic or charge density wave fluctuations. Since the cyclotron orbit in momentum space has to pass through these hot spots, and the scattering of the electrons at the hot spots is not forward in nature due to the finite boson wavevector, a small amount of cyclotron transport mass renormalization is to be expected. A detailed study is, however, quite complicated to perform using the Landau level basis methods developed in Ref. \cite{Aldape2022} and this work, and is therefore deferred for future work.

\acknowledgments We thank Ilya Esterlis for discussion and collaboration on related work. We thank Peter Armitage for reading our draft and providing valuable comments. This work was partially performed at the Aspen Center for Physics, which is supported by National Science Foundation grant PHY-2210452. The participation of H.G. at the Aspen Center for Physics was supported by the Simons Foundation. This research was partially conducted at the Kavli Institute for Theoretical Physics, supported by the National Science Foundation under Grants No. NSF PHY-1748958 and PHY-2309135. This research was also supported by the U.S. National Science Foundation grant No. DMR-2245246 and by the Simons Collaboration on Ultra-Quantum Matter which is a grant from the Simons Foundation (651440, S.S.). The Flatiron Institute is a division of the Simons Foundation. \joerg{J.S. and D. V. were supported by the German Research Foundation (DFG) through CRC TRR 288 “ElastoQMat,” project A07.}

\bibliography{fermi2,fermi3}
\end{document}